\documentclass[epjST,amsfonts]{svjour}
\usepackage{graphicx}
\usepackage{latexsym}
\usepackage{bm}
\usepackage{amsmath}
\begin{document}
\newcommand{\leftg}{\langle \phi_0 |}
\newcommand{\rightg}{| \phi_0 \rangle}
\newcommand{\vect}[1]{\stackrel{\rightarrow}{#1}}
\newcommand{\rvec}{\vec{r}}
\newcommand{\nn}{\nonumber}
\newcommand{\sigmavec}{\boldsymbol{\mathbf\sigma}}
\newcommand{\tauvec}{\boldsymbol{\mathbf\tau}}
\title{Nuclear Energy Density Functionals Constrained by Low-Energy QCD}
\author{Dario Vretenar\inst{1}\fnmsep\thanks{Work supported in part
by MZOS - 
project 1191005-1010, \email{vretenar@phy.hr}} }
\institute{Physics Department, University of Zagreb, 
Bijeni\v cka 32, Zagreb, Croatia}
\abstract{A microscopic framework of nuclear energy density functionals 
is reviewed, which establishes a direct relation between low-energy QCD and
nuclear structure, synthesizing effective field theory 
methods and principles of density functional theory.
Guided by two closely related features of QCD in the 
low-energy limit: a) in-medium changes of vacuum condensates, and 
b) spontaneous  breaking of chiral symmetry; a relativistic energy 
density functional is developed and applied in studies of ground-state  
properties of spherical and deformed nuclei.
} 
\maketitle
\section{Introduction}
\label{intro}
Modern nuclear structure is rapidly evolving from studies of nuclei 
close to the $\beta$-stability line towards regions of exotic short-lived 
nuclei and systems at the nucleon drip-lines. A wealth 
of new experimental results has also prompted important qualitative and 
quantitative advances in nuclear structure theory. The 
principal objective is to build a consistent microscopic framework that will 
provide a unified description of bulk properties, 
nuclear excitations and reactions. 

Among the microscopic approaches to the nuclear many-body problem, the 
framework of nuclear energy density functionals (NEDF) is the only one that 
can be applied over the whole chart of nuclides, from relatively light systems 
to superheavy nuclei, and from the valley of $\beta$-stability to the particle 
drip-lines. NEDFs are not necessarily related to any realistic nucleon-nucleon 
interaction, but rather represent global functionals of nucleon densities and  
currents. With a small set of universal parameters adjusted to data, EDFs 
have achieved a high level of accuracy in the description of structure properties. 
Nevertheless, one expects that new data on exotic nuclei with extreme isospin 
values will present interesting challenges for the NEDF framework.

Even though completely phenomenological EDFs have been successfully 
employed in extensive structure studies, it would clearly be desirable to have
a fully microscopic foundation for a universal EDF framework. Eventually, 
the description of the nuclear many-body problem, including both extended 
nuclear matter and finite nuclei, must be linked to and constrained by the 
underlying theory of strong interactions. In these lectures we will review a novel 
program \cite{Fi.02,Fi.03,Fi.06,Fi.07}, 
which integrates effective field theory (EFT) methods and principles of density 
functional theory (DFT), to bridge the gap between features of low-energy 
non-perturbative quantum chromodynamics (QCD) and nuclear phenomenology. 
Guided by two important elements that establish
connections with chiral dynamics and the symmetry breaking pattern 
of low-energy QCD: a) strong scalar and vector nucleon fields related to 
in-medium changes of QCD vacuum condensates; 
b) the long- and intermediate-range interactions generated by one-and 
two-pion exchange, as derived from in-medium chiral perturbation theory; 
a relativistic nuclear EDF has been developed which includes second-order 
gradient corrections to the local density approximation. Translated into 
a relativistic point-coupling model with density-dependent vertices, 
this framework has been applied in calculations of ground-state properties 
over a broad range of spherical and deformed nuclei.

Sec.~\ref{sec:DFT} introduces elementary concepts of density functional 
theory, and in Sec.~\ref{sec:NucEDF} we discuss some basic issues in the 
construction of nuclear EDFs. Elements of low-energy QCD are reviewed 
in Sec.~\ref{sec:QCD}. In-medium chiral EFT is applied in the calculation 
of the nuclear matter equation of state in Sec.~\ref{sec:NM}. In 
Sec.~\ref{sec:FKVW} we develop a relativistic EDF based on chiral 
pion-nucleon dynamics, and present some illustrative calculations 
of nuclear ground-state properties. A brief summary is included in 
Sec.~\ref{sec:Conclusion}.

\section{Basic Concepts of Density Functional Theory}
\label{sec:DFT}

Density Functional Theory (DFT) is one of the most popular and successful 
``ab initio" approaches to the structure of quantum many-body systems (atoms, molecules, solids).  Probably no other method achieves comparable 
accuracy at the same computational cost. The basic concept is that the ground-state
properties of a stationary many-body system can be represented in terms 
of the ground-state density alone. Since the density $n(\vec{r})$ is a function of 
only three spatial coordinates, rather than the 3N coordinates of the N-body 
wave function, DFT is computationally feasible even for large systems. 

In this section we introduce elementary concepts of DFT, that will also be useful in 
considering nuclear energy density functionals. Detailed expositions are available 
in numerous review articles, textbooks and collection of research papers. 
We refer the reader, for instance, to \cite{Kohn.99,AM.00,DG.90,PY.89,FNM.03}.
The basis of DFT are the Hohenberg-Kohn theorem and the Kohn-Sham scheme, 
which will be introduced in the next two sections. We will also briefly discuss the 
approximations to the central ingredient of the DFT approach: the exchange-correlation energy functional. 
\subsection{The Hohenberg-Kohn Theorem}
\label{sec:HK}
In ground-state DFT one considers a system of N interacting particles 
described by the Hamiltonian:
\begin{equation}
\hat{H} \equiv \hat{T} + \hat{V} + \hat{W} =
- \sum_{i=1}^{N} {{\nabla_i^2}\over{2m}} 
+ \sum_{i=1}^{N} v({\vec r}_i)  + 
{1\over 2} \sum_{i=1}^{N}  \sum_{j\neq i}^{N} w({\vec r}_i,{\vec r}_j) \; .
\end{equation}
Let $\Psi$ be the N-body wave function and $n({\vec r})$
the corresponding particle 
density:
\begin{equation}
n({\vec r}) = N \int d^3r_2 \int d^3r_3 \ldots \int d^3r_N ~
\Psi^*({\vec r},{\vec r_2},\ldots ,{\vec r_N})
\Psi({\vec r},{\vec r_2},\ldots ,{\vec r_N}) \; .
\end{equation}
The Hohenberg-Kohn (HK) theorem \cite{HK.64} can be 
summarized in the following statements:
\begin{itemize}
\item
The nondegenerate ground-state (GS) wave function is a unique functional 
of the GS density:
\begin{equation}
\Psi_0({\vec r_1},{\vec r_2},\ldots ,{\vec r_N}) = \Psi[n_0({\vec r})] \; .
\end{equation}
As a consequence, the GS expectation value of any observable ${\cal O}$ 
is also a functional of $n_0({\vec r})$:
\begin{equation}
{\cal O}_0 \equiv {\cal O}[n_0] = \langle \Psi [n_0] | \hat {\cal O} | \Psi [n_0] \rangle \; .
\end{equation}
\item The GS energy:
\begin{equation}
E_0 \equiv E[n_0] =   \langle \Psi [n_0] | \hat { H} | \Psi [n_0] \rangle \; , 
\end{equation}
and the GS density  $n_0({\vec r})$ of a system characterized by the 
external potential $v_0({\vec r})$ can be obtained from a variational 
principle which involves only the density, i.e. the inequality:
\begin{equation}
E_0 = E_{v_0}[n_0] < E_{v_0}[n] \; ,
\end{equation}
holds for all other densities $n({\vec r})$.
\item There exists a functional $F[n]$ such that the energy functional can 
be written as: 
\begin{equation}
E_{v_0}[n] = F[n] + \int d^3r~v_0({\vec r})n({\vec r}) \; .
\end{equation}
The functional $F[n]$ is {\em universal} in the sense that, for a given 
particle-particle interaction (for instance the Coulomb interaction), 
it does not depend on the potential $v_0({\vec r})$, i.e. it has the same 
functional form for all systems.
\end{itemize}
The proof is based on the Rayleigh-Ritz variational principle \cite{Kohn.99,DG.90,HK.64}.
The formal definition of the Hohenberg-Kohn functional $F[n]$ then reads:
\begin{equation}
F[n] = T[n] + W[n] = \langle \Psi [n] | \hat {T} | \Psi [n] \rangle +
\langle \Psi [n] | \hat {W} | \Psi [n] \rangle \; ,
\end{equation}
where $ \Psi [n]$ is the N-body wave function which minimizes the expectation 
value of $\hat {T} + \hat {W}$. The GS density uniquely determines the 
external potential of the system and thus, as the kinetic energy and 
particle-particle interaction are universal functionals, the entire Hamiltonian 
and all physical properties of the system. However, although it establishes 
the variational character of the energy functional, the HK theorem
gives no practical guide to the construction of the universal functional, and 
the explicit density dependence of $F[n]$ is not known. In order to make 
practical use of DFT, one needs reliable approximations for $T[n]$ and $W[n]$.

\subsection{The Kohn-Sham Scheme}
\label{sec:KS}
Most practical applications of DFT use the effective single-particle 
Kohn-Sham (KS) equations, 
which are introduced for an auxiliary system of N non-interacting 
particles described by the Hamiltonian:
\begin{equation}
\hat{H}_s = \hat{T} + \hat{V}_s \; .
\end{equation}
The HK theorem states that there exists a unique energy functional 
\begin{equation}
E_s[n] = T_s[n] + \int d^3r~v_s({\vec r}) n({\vec r}) \; ,
\end{equation}
for which the variational equation yields the exact ground-state density 
$n_s({\vec r})$ corresponding to $\hat{H}_s$, and $T_s[n]$ is 
the universal kinetic energy functional of the non-interacting system. 
The KS scheme is based on the following assertion: 
for any interacting system, there exists a unique local 
single-particle potential $v_s({\vec r})$, such that the exact GS density of the 
interacting system equals the GS density of the auxiliary non-interacting system:
\begin{equation}
n({\vec r})=n_s({\vec r}) = \sum_i^{N} |\phi_i ({\vec r})|^2 \; ,
\label{occ}
\end{equation}
expressed in terms of the lowest N single-particle orbitals -- solutions of the 
Kohn-Sham equations:
\begin{equation}
\left [ - {{\nabla^2}/{2m}} + v_s({\vec r}) \right ] \phi_i ({\vec r}) 
= \varepsilon_i \phi_i ({\vec r}) \; .
\label{KS_eq}
\end{equation}
The uniqueness of $v_s({\vec r})$ follows from the HK theorem, and the 
single-particle orbitals are unique functionals of the density: 
$\phi_i ({\vec r}) = \phi_i ([n];{\vec r})$. The HK functional is partitioned 
in the following way:
\begin{equation}
F[n] = T_s[n] + U[n] + E_{xc}[n] \; ,
\label{xc}
\end{equation}
where $U[n]$ is the Hartree functional: 
\begin{equation}
U[n] = {1\over 2} \int d^3 r \int d^3 r' n({\vec r}) w({\vec r}, {\vec r'})n({\vec r'}) \; ,
\end{equation}
and the last term is the {\em exchange-correlation} energy functional 
$E_{xc}[n]$  which, by the definition Eq.~(\ref{xc}), includes everything else -- 
all the many-body effects. It can formally be defined by:
\begin{equation}
E_{xc}[n] = T[n] + W[n] - U[n] - T_s[n] \; .
\end{equation} 
The corresponding effective Kohn-Sham potential therefore reads:
\begin{equation}
v_s[n]({\vec r}) = v({\vec r}) + \int d^3 r'  w({\vec r}, {\vec r'})n({\vec r'}) +
v_{xc}[n]({\vec r}) \; ,
\label{v_KS}
\end{equation}
where the local {exchange-correlation potential} is defined by:
\begin{equation}
v_{xc}[n]({\vec r})  = {{\delta E_{xc}[n]}\over {\delta n({\vec r})}} \; .
\end{equation}
Since the effective potential depends on the density, the system of equations 
(\ref{occ}), (\ref{KS_eq}), and (\ref{v_KS}) has to be solved self-consistently. 
This is the Kohn-Sham scheme of density functional theory \cite{KS.65}. 

By construction, therefore, the exact $v_s[n_o]({\vec r})$ is the unique external 
potential which, for the non-interacting system, leads to the same physical 
GS density $n_0({\vec r})$ as that of N interacting particles in the external 
potential $v({\vec r})$. In the limit when one neglects $E_{xc}$ ($v_{xc}$), 
the KS equations reduce to the self-consistent Hartree equations. However, 
since it includes correlation effects, the full KS scheme goes beyond the 
Hartree-Fock approximation and, in addition, it has the advantage of being 
a {\em local} scheme. With an {\em exact} functional  
$E_{xc}$, all many-body effects are in principle included. 
Thus the usefulness of the Kohn-Sham scheme crucially depends 
on our ability to construct accurate approximations to the 
exact exchange-correlation energy. 
\subsection{Approximations for the Exchange-Correlation Energy}
\label{sec:EXC}
The true exchange-correlation energy functional is {\em universal}, 
i.e. given the inter-particle interaction, it has the same functional form for 
all systems. The construction of an expression for the unknown functional 
$E_{xc}$ is therefore the keystone of the Kohn-Sham 
DFT. One possible approach is to develop $E_{xc}$ from first principles 
by incorporating known exact constraints. Another is empirical, 
a parametric ansatz is optimized by adjusting it to a set of data. 
Modern approximations for $E_{xc}$ typically combine
both strategies.

The simplest, but at the same time still useful, is the well known local 
density approximation (LDA). In LDA, the exchange-correlation energy 
functional reads:
\begin{equation}
E_{xc}^{LDA} [n] = \int d^3r~n({\vec r}) e_{xc}^{unif}(n({\vec r})) \; ,
\end{equation} 
where $e_{xc}^{unif}(n({\vec r}))$ is the exchange-correlation energy per 
particle of a uniform infinite system of density $n$ (electron gas, nuclear 
matter), and which can be obtained from an ``ab initio" calculation. By its construction, 
the LDA is {\em a priori} expected to be valid for spatially slowly varying 
densities. Nevertheless, LDA proved to be a surprisingly accurate approximation 
also for systems with rapid density variations. This is a consequence of exact 
properties which the LDA inherits from the homogeneous system. 

Any realistic system is spatially inhomogeneous, and it is clearly useful to 
incorporate the information on the rate of density variation in $E_{xc}$.
In the generalized-gradient approximation (GGA) the  exchange-correlation 
functional is written as:
\begin{equation}
E_{xc}^{GGA} [n] = \int d^3r~f(n({\vec r}), \nabla n({\vec r})) \; .
\end{equation} 
GGA is more accurate then LDA but, unlike the input $e_{xc}^{unif}$ in 
LDA, the function $f$ is not unique and, depending on the method of constructing 
$f(n({\vec r}), \nabla n({\vec r}))$, very different GGAs can be obtained. For 
instance, while typical GGAs used in quantum chemistry \cite{LYP.88} 
are based on parametric 
forms fitted to sets of selected molecules, exact constraints are incorporated 
in first-principles GGAs \cite{PBE.96} used in most physics applications.

The next rung on the ladder of approximations for $E_{xc}$ are ``meta-GGA" 
functionals which, in addition to the density and its gradient, depend on the 
kinetic energy density of the occupied Kohn-Sham orbitals
\begin{equation}
\tau({\vec r}) = {1\over 2} \sum_i^{occ} |\nabla \phi_i ({\vec r})|^2 \; .
\end{equation}
Unlike LDA and GGA, which are explicit functionals of the density alone, 
the meta-GGA functional 
\begin{equation}
E_{xc}^{MGGA} [n] = \int d^3r~g(n({\vec r}), \nabla n({\vec r}), \tau({\vec r})) 
\end{equation} 
also depends explicitly on the Kohn-Sham orbitals. Meta-GGA is the highest 
level of approximation which does not include full non-locality. Approximations 
of even higher level of accuracy incorporate increasingly complex ingredients, 
and include fully non-local functionals of the Kohn-Sham orbitals, occupied as
well as unoccupied.
%
\section{Nuclear energy density functionals}
\label{sec:NucEDF}
%

The most complete and accurate description of structure phenomena 
in heavy nuclei is currently provided by self-consistent mean-field (SCMF) models.
A variety of ground-state properties and collective excitations, not only in 
medium-heavy and heavy stable nuclei, but also in regions of exotic 
nuclei far from the line of $\beta$-stability and close to the nucleon drip-lines, 
have been successfully described with mean-field models based on the Gogny interaction, the Skyrme energy functional, and the relativistic meson-exchange 
effective Lagrangian \cite{BHR.03,VALR.05}. In the mean-field approximation 
the dynamics of the nuclear many-body system is represented by independent 
nucleons moving in a self-consistent potential, which corresponds to the actual 
density distribution of a given nucleus. Important advantages of using 
the mean-field framework include the use of global effective nuclear interactions, 
the treatment of arbitrarily heavy systems including superheavy nuclei, and the 
intuitive picture of intrinsic shapes. 

The SCMF approach to nuclear structure is analogous to Kohn-Sham density 
functional theory. As we have shown in the previous section, DFT
enables a description of a quantum many-body problem in terms of
a universal energy density functional, and nuclear mean-field models
approximate the exact energy functional, which includes all
higher-order correlations, with powers and gradients of
ground-state nucleon densities and currents. Although it models the effective
in-medium interaction between nucleons, a general density functional 
is not necessarily related to any given NN potential.
By employing global parameter sets, adjusted to reproduce empirical 
properties of symmetric and asymmetric nuclear matter, and bulk properties
of simple, spherical and stable nuclei, the current generation of 
SCMF models has been applied in numerous studies of 
structure phenomena over the whole nuclear chart.

The earliest applications of DFT in nuclear structure \cite{VB.72,Neg.70,NV.72} 
used the zero-range density-dependent effective Skyrme interaction. 
The corresponding Skyrme functional can be written as 
the most general energy-density functional in isoscalar and isovector 
density, spin density, current, spin-current tensor, kinetic
density, and kinetic spin density, respectively \cite{BHR.03}:
\begin{eqnarray}
&\rho_0 (\rvec)
  =   \sum_{\sigma \tau} \rho(\rvec\sigma \tau;\rvec\sigma \tau) 
      \quad \quad
&\rho_1 (\rvec)
  =   \sum_{\sigma \tau} \rho(\rvec\sigma \tau;\rvec\sigma \tau) \; \tau
      \nn \\
&\vec{s}_0 (\rvec)
  =   \sum_{\sigma\sigma'\tau} \rho(\rvec\sigma \tau;\rvec\sigma'\tau) \,
      \sigmavec_{\sigma' \sigma} 
      \quad \quad
&\vec{s}_1 (\rvec)
 =   \sum_{\sigma\sigma'\tau} \rho(\rvec\sigma \tau;\rvec\sigma'\tau) \,
      \sigmavec_{\sigma' \sigma} \; \tau
      \nn \\
&\vec{j}_T(\rvec) =  \frac{i}{2} (\nabla' - \nabla) \, 
      \rho_T(\rvec, \rvec')\big|_{\rvec=\rvec'} 
      \quad \quad
&{\cal{J}}_T(\rvec)
=  \frac{i}{2} (\nabla' - \nabla) \otimes  
      \vec{s}_T(\rvec, \rvec')\big|_{\rvec=\rvec'} 
      \nn \\
&\tau_T(\rvec)
 = \nabla \cdot \nabla' \, \rho_T (\rvec, \rvec') \big|_{\rvec=\rvec'} 
 \quad \quad   
&\vec{T}_T(\rvec)
 =   \nabla \cdot \nabla' \, 
       \vec{s}_T (\rvec, \rvec') \big|_{\rvec=\rvec'} \; ,
\label{eq:locdensities}
\end{eqnarray}
where $\sigma$ denotes the spin, and $\tau$ the isospin of the nucleon.
Isoscalar ({$T=0$}) quantities are sums 
({$\rho_0 = \rho_n + \rho_p$}), whereas isovector 
($T=1$) densities correspond to proton-neutron differences
({$\rho_1 = \rho_n - \rho_p$}). 
The Skyrme functional contains systematically all 
possible bilinear terms in the local densities and currents of 
Eq.~(\ref{eq:locdensities}) up to second order in the derivatives, which
are invariant with respect to parity, time-reversal, rotational,
translational and isospin transformations:
\begin{eqnarray}
{E}_{\rm Sk}
& = & \sum_{T = 0,1} \left \{~
C_T^{\rho} \, \rho_{T}^{2}
      + C_T^{\Delta \rho} \, \rho_{T} \Delta \rho_{T}
      + C_T^{\tau} \, \rho_{T} \tau_{T}
      + C_T^{J}{\cal{J}}^2_{T} 
      + C_T^{\nabla J} \rho_{T} \, \nabla\!\cdot\!\vec{J}_{T} \right.
      \nn \\
&   & 
 \left. C_T^{s} \, \vec{s}^{2}_{T}
      + C_T^{\Delta s} \, \vec{s}_{T}\!\cdot\!\Delta \vec{s}_{T}
      + C_T^{sT} \, \vec{s}_{T} \!\cdot\! \vec{T}_{T}      
      + C_T^{\nabla s} \, (\nabla \!\cdot\! \vec{s}_T)^2  
      + C_T^{j} \, \vec{j}^2_{T}
      + C_T^{\nabla j} \, 
        \vec{s}_{T} \!\cdot\! \nabla\!\times\! \vec{j}_{T} ~\right \}
\label{E_Sk}
\end{eqnarray}
where the coefficients $C_T$ in the isoscalar and isovector channels can be either 
constants or explicitly depend on the nucleon density.

Even though the functional can be derived from the ground-state expectation 
value of a zero-range momentum-dependent force introduced by Skyrme, 
in modern applications the functional is parameterized directly by fitting  
the coefficients to nuclear ground-state data, without reference to any 
NN interaction. Over the last thirty years more than hundred different Skyrme parameterizations have been adjusted and analyzed, and about 
ten of these parameter sets are still used in extensive studies of nuclear 
structure phenomena. This means, however, that often it is difficult to compare 
results obtained with different models, also because they include different 
subsets of terms from the general functional Eq.~(\ref{E_Sk}). Ideally, 
model dependences could be removed by including all terms allowed by 
symmetries \cite{Per.04}. However, it seems that available data can only 
constrain a subset of 
parameters, and one needs additional criteria for selecting the optimal 
energy density functional form.

One of the major goals of modern theoretical nuclear physics is, therefore, 
to build a universal nuclear energy density functional (NEDF) \cite{LNP.641}. 
Universal in the sense that the same functional form is used for all
nuclei, with the same set of parameters. 
This framework should then provide a reliable microscopic description of 
infinite nuclear and neutron matter, ground-state properties of 
all bound nuclei, low-energy vibrations, rotational spectra, 
small-amplitude vibrations, and large-amplitude adiabatic properties.   
In order to formulate a microscopic NEDF, however, one must be able to go
 beyond the mean-field approximation and systematically calculate the exchange-correlation energy functional $E_{xc}[\rho]$. Many-body correlations must be 
included  starting from the relevant active degrees of freedom at low-energies 
characteristic for nuclear binding.  

Nuclei are aggregates of nucleons (and mesons), which are in turn clusters
of quarks and gluons. The decription of nuclear many-body systems must 
therefore ultimately be linked to and constrained by the underlying theory 
of strong interactions -- quantum chromodynamics (QCD).
The success of the non-relativistic and covariant SCMF approach to 
nuclear many-body dynamics \cite{BHR.03,VALR.05}, and the recent 
application of chiral effective field theory to nucleon-nucleon scattering and the 
few-body problem \cite{Epel.06}, point to a unified microscopic framework
based on density functional theory (DFT) and effective 
field theories (EFT). As we will show in the following sections, within this framework 
a NEDF can be developed in a systematic way by using an EFT of low-energy 
in-medium nucleon-nucleon interactions to construct accurate approximations 
to the exact exchange-correlation energy functional.

Effective field theories are low-energy approximations to more fundamental 
theories. Instead of trying to solve explicitly the underlying theory, low-energy 
processes are described in terms of active degrees of freedom. The low-energy 
behavior of the underlying theory must be known, and it must be built into the
EFT. High-momentum states that characterize the unknown short-distance 
dynamics must be excluded, and the EFT is used to calculate observables in 
terms of an expansion in $p/\Lambda$ ($p$ denotes momenta or masses 
smaller than a certain momentum scale $\Lambda$). The resulting effective 
Lagrangian must include all terms that are compatible with the symmetries of 
the underlying theory and this means, in principle, an infinite number of terms. 
A working EFT must, therefore, be completed by specifying a method that allows 
to decide which terms of the effective Lagrangian contribute in a calculation up to a
desired level of accuracy. 

In the next section we will outline basic concepts of chiral EFT, a low-energy 
approximation to QCD. At low energies characteristic for nuclei, the active degrees 
of freedom of QCD are pions and nucleons, and their dynamics is governed 
by the chiral $SU(2) \times SU(2)$ symmetry and its spontaneous breakdown.
%
\section{Elements of Low-Energy QCD}
\label{sec:QCD}
%
Quantum chromodynamics is the fundamental quantum field theory of strong interactions. 
At very short distance scales, $r < 0.1$ fm,  corresponding to momentum transfers
exceeding several GeV/c which probe space-time intervals deep inside the nucleon, 
QCD is a theory of weakly interacting quarks and gluons. At momentum scales 
considerably smaller than 1 GeV (corresponding to length scales $r \geq 1$ fm,
typical for the physics of atomic nuclei), QCD is characterized
by color confinement and a non-trivial vacuum: the ground state of QCD hosts 
strong condensates of quark-antiquark pairs and gluons. The associated spontaneous
breaking of chiral symmetry implies the existence of pseudoscalar Goldstone bosons. 
For two quark flavours  ($u$ and $d$ quarks), the Goldstone bosons are identified with 
the isospin triplet of pions. At low energy, pions interact weakly with one another and 
with any massive hadron. At the energy scale of nuclear binding, QCD is thus 
realized as an effective field theory of pions coupled to nucleons.

The quark condensate $\langle \bar{q}q \rangle$, i.e. the ground state expectation value of the scalar quark density, plays a particularly important role as an order parameter of spontaneously broken chiral symmetry. Hadrons and nuclei represent excitations built on this condensed ground state. Changes of the condensate structure of the QCD vacuum in the presence of baryonic matter are a source of strong fields felt by the nucleons which, together with the pion-exchange forces, govern nuclear dynamics. 

In this section we briefly outline some basic concepts of QCD in the low-energy 
two-flavour ($N_f = 2$) sector. For a more detailed introduction we refer the reader to 
Refs.~\cite{PS.95,Mut.98,TW}.
\subsection{The QCD Lagrangian}
\label{sec:QCD-L}
In the low-energy sector of QCD the lightest $u$ and $d$ quarks form a flavour $N_f = 2$ (isospin) doublet with "bare" quark masses of less than 10 MeV. The flavour (and colour) components of the quarks are collected in the Dirac fields $\psi (x) = (u(x), d(x))^T$. The QCD Lagrangian
\begin{equation}
{\cal L}_{QCD} = \bar{\psi} (i \gamma_{\mu} D^{\mu} - m) \psi - \frac{1}{2} Tr
(G_{\mu \nu} G^{\mu \nu}),
\label{QCD}
\end{equation}
includes the gluonic field strength tensor
\begin{equation}
G_{\mu\nu} = \partial_{\mu} A_{\mu}^a  t_a - \partial_{\nu} A_{\mu}^a t_a 
+ g f_{abc} A_{\mu}^a A_{\nu}^b  t_c \; ,
\end{equation}
 the covariant derivative
\begin{equation}
D_{\mu} = \partial_{\mu} - i g A_{\mu}^a t_a \; ,
\end{equation}
and the generators $ t_a  \equiv \lambda_a /2$ ($a=1,\ldots,8$)
of $SU(3)_{\rm color}$ group.
The $2 \times 2$ matrix $m = diag (m_u, \, m_d)$ contains the light quark masses. 
Since quarks are not observed as asymptotically free states, 
the quark mass depends on the momentum scale at which it is probed. At a scale of 
the order of 1 GeV: $m_u = 4 \pm 2$ MeV,  $m_d = 8 \pm 4$ MeV \cite{PP}. 
The strange quark is more than an order of magnitude heavier ($m_s \sim$ 150 MeV), 
and it will not be considered here. 

The interaction between quarks and gluons is independent of the quark flavour. 
The coupling strength $\alpha_s = g^2 /4\pi$ depends on 
the momentum scale $Q$. At high energies,  $\alpha_s$ is small and QCD is characterized
by  asymptotic freedom. In this regime it is possible to apply perturbation theory 
with an expansion in $\alpha_s$.  For $Q \leq 1$ GeV,  $\alpha_s$ increases sharply
and the relevant degrees of freedom are no longer the elementary colored quarks and gluons. 
They are replaced by color neutral bound states: the hadrons. Color confinement characterizes 
QCD at low energies. 
\subsection{Chiral Symmetry}
\label{sec:QCD-ChS}
Let us consider QCD in the limit of massless quarks, by setting $m=0$ in Eq.(\ref{QCD}). In this limit, the QCD Lagrangian has a global symmetry related to the conserved right- or left-handedness (chirality) of zero mass spin $1/2$ particles. Introducing right- and left-handed quark fields,
\begin{equation}
\psi_{R,L} = \frac{1}{2} (1 \pm \gamma_5) \psi ,
\end{equation}
we observe that separate global unitary transformations
\begin{equation}
\psi_R \to \exp [i \theta^a_R \frac{\tau_a}{2}] \, \psi_R, \hspace{1,5cm} \psi_L
\to \exp [i \theta^a_L \frac{\tau_a}{2}] \, \psi_L ,
\end{equation}
with $\tau_a (a= 1,2,3)$ the generators of (isospin) $SU(2)$, leave ${\cal L}_{QCD}$ invariant in the limit $m \to 0$ 
\begin{equation}
{\cal L}_{QCD} \to  \bar{\psi}_L i \gamma_{\mu} D^{\mu} \psi_L 
+  \bar{\psi}_R i \gamma_{\mu} D^{\mu} \psi_R
\end{equation}
This is the chiral $SU(2)_R \times SU(2)_L$ symmetry of QCD. It implies six conserved Noether currents,
$J^{\mu}_{R, a} = \bar{\psi}_R \gamma^{\mu} {\tau_a}/{2} \psi_R$ and
$J^{\mu}_{L, a} = \bar{\psi}_L \gamma^{\mu} {\tau_a}/{2} \psi_L$, with
$\partial_{\mu} J^{\mu}_R = \partial_{\mu} J^{\mu}_L = 0 $. It is also useful to
introduce the vector and axial currents,
\begin{equation}
V^{\mu}_a = J^{\mu}_{R,a} + J^{\mu}_{L,a} = \bar{\psi} \gamma^{\mu }\frac{\tau_a}{2} \psi  \, , \hspace{7mm}
A^{\mu}_a (x) = J_{R,a} - J_{L,a} = \bar{\psi} \gamma^{\mu} \gamma_5 \frac{\tau_a}{2} \psi\,  .
\end{equation}
Their corresponding charges,
\begin{equation}
Q^V_a = \int d^3 x 
~\psi^{\dagger} (x) \frac{\tau_a}{2} \psi (x)\, , \hspace{1cm}
Q^A_a = \int d^3 x 
~\psi^{\dagger} (x) \gamma_5 \frac{\tau_a}{2} \psi (x) \, ,
\label{charge}
\end{equation}
are, likewise, generators of $SU(2) \times SU(2)$.

Chiral symmetry is explicitly broken by the finite quark masses, 
because the mass term in the Lagrangian mixes left- and right-handed fields: 
\begin{equation}
{\cal L}_{m} = - \bar{\psi} m  \psi =  -(\bar{\psi}_R m  \psi_L 
+  \bar{\psi}_L m  \psi_R) \; .
\end{equation}
As a consequence, the charge operators Eq.~(\ref{charge}) are, in general, no longer
time independent.
\subsection{Spontaneous Chiral Symmetry Breaking}
\label{sec:QCD-SCSB}
A (continuous) symmetry is said to be spontaneously broken or 
hidden,  if the ground state of the system is not invariant under the full 
symmetry of the Lagrangian. In the case of the QCD Lagrangian with $m = 0$, 
the chiral $SU(2) \times SU(2)$ symmetry 
is spontaneously broken:  the ground state (vacuum) of QCD has lost part 
of the symmetry of the Lagrangian. It is symmetric only under the 
subgroup $SU(2)_V$ generated by the vector charges $Q^V$ (isospin 
symmetry). 

Evidence for the spontaneous breaking of chiral symmetry is found in 
the spectrum of physical hadrons. If the QCD ground state were symmetric under 
chiral $SU(2) \times SU(2)$,  it would be annihilated both by the vector and axial 
charge operators: $Q^V_a |0 \rangle = Q^A_a |0 \rangle = 0$. This would then imply 
the existence of parity doublets in the hadron spectrum which, however, are 
not observed in nature. For instance, the vector $(J^{\pi} = 1^-)$ $\rho$ meson mass ($m_{\rho} \simeq 0.77$ GeV) is well separated from that of the axial vector 
$(J^{\pi} = 1^+)$ $a_1$ meson ($m_{a_1} \simeq 1.23$ GeV). Likewise, the light 
pseudoscalar $(J^{\pi} = 0^-)$ mesons have masses much lower than the lightest scalar 
$(J^{\pi} = 0^+)$ mesons.

Goldstone's theorem states that in a quantum field theory every spontaneously 
broken continuous global symmetry leads to massless particles (Goldstone bosons) 
with the same quantum numbers as the generators of the broken symmetry. 
If G is the symmetry group of the Lagrangian with $n_G$ generators, and 
H the subgroup with $n_H$ generators which leaves the ground state invariant 
after spontaneous symmetry breaking, the total number of Goldstone bosons 
equals  $n_G - n_H$.

In the case of chiral symmetry, to each axial 
generator $Q^A_a$ of $SU(2)_A$, which does 
not annihilate the ground state, corresponds a massless Goldstone boson field 
with spin $0$ and negative parity. Namely, if $Q^A_a | 0 \rangle \neq 0$, there
must be a physical state generated by the axial charge, 
$|\phi_a \rangle = Q^A_a | 0 \rangle$, which is
energetically degenerate with the vacuum. Let $H_0$ be the QCD Hamiltonian
(with massless quarks) which commutes with the axial charge. Setting the ground
state energy equal to zero for convenience, we have $H_0 |\phi_a \rangle = Q^A_a
H_0 | 0 \rangle = 0$. Evidently $| \phi_a \rangle $ represents three massless
pseudoscalar bosons (for $N_f = 2$). They are identified with the pions:
$\pi^-$, $\pi^0$, and $\pi^+$.

Goldstone's theorem consequently implies non-vanishing matrix elements of the 
axial current between the vacuum and the pion $\langle 0 | A^{\mu} | \pi \rangle$, 
which must be proportional to the pion momentum:
\begin{equation}
\langle 0 | A^{\mu}_a (x) | \pi_b (q)\rangle = - i f_\pi q^\mu \delta_{ab} e^{-i q \cdot x} \; ,
\label{pi_decay}
\end{equation}
and the constant of proportionality is the pion decay constant 
$f_{\pi} = 92.4$ MeV, determined from the decay  
$\pi^+ \rightarrow \mu^+ \nu_\mu +  \mu^+ \nu_\mu \gamma$.  The divergence of 
Eq.~(\ref{pi_decay}) reads:
\begin{equation}
\langle 0 | \partial_\mu A^{\mu}_a (x) | \pi_b (q)\rangle = - f_\pi q^2 \delta_{ab} e^{-i q \cdot x} 
=  - f_\pi m_\pi^2 \delta_{ab} e^{-i q \cdot x} \; .
\label{PCAC}
\end{equation}
The small pion mass ($\approx 140$ MeV), as compared to hadronic scales, 
is directly linked to the partial conservation of the axial current 
(PCAC) Eq.~(\ref{PCAC}). The axial transformation is thus an approximate symmetry 
of QCD, and becomes exact in the limit $ m \to 0$.
\subsection{The Chiral Condensate}
\label{sec:QCD-ChCond}
Spontaneous symmetry breaking is related to a scalar operator whose 
nonvanishing vacuum expectation value plays the role of the order parameter
of the symmetry breaking. The QCD ground state contains scalar quark-antiquark 
pairs, and the corresponding expectation value $\langle 0 | \bar{\psi} \psi | 0 \rangle$ is called the chiral (or quark) condensate:
\begin{equation}
\langle \bar{\psi} \psi \rangle \equiv \langle \bar{u} u \rangle + \langle \bar{d} d
\rangle =  - {\rm Tr} \lim_{y \to x^+} \langle 0 | {\cal T} \psi (x)
\bar{\psi} (y) | 0 \rangle \; ,
\end{equation}
where ${\cal T}$ denotes the time-ordered
product.  The relation between spontaneous chiral symmetry 
breaking and the non-vanishing chiral condensate can be derived from 
the (equal-time) commutator between 
the pseudoscalar operator
$P_a (x) = \bar{\psi} (x) \gamma_5 \tau_a \psi (x)$ and 
 the axial charge $Q^A_a$ of eq.(\ref{charge}): 
$[ Q^A_a, P_b] = - \delta_{ab} \bar{\psi} \psi$.
Taking the ground state expectation value of this commutator, 
we notice that $Q^A_a | 0 \rangle \neq 0$ is consistent 
with $\langle \bar{\psi} \psi \rangle \neq 0$.

The Goldstone boson (pion) acquires its physical mass 
through the explicit breaking of chiral symmetry by the finite quark masses 
$m_{u,d}$. The pion mass $m_{\pi}$ is related to the $u$- and $d$- quark 
masses by the Gell-Mann, Oakes, Renner (GOR) relation \cite{GOR}:
\begin{equation}
m^2_{\pi} = - \frac{1}{f^2} (m_u + m_d) \langle \bar{q} q \rangle + 
{\cal O}(m^2_{u,d}) .
\label{GOR}
\end{equation}
Isospin symmetry ($Q^V_a | 0 \rangle = 0$) has been used 
in the definition:
$\langle \bar{q} q \rangle \equiv \langle \bar{u} u \rangle \simeq
\langle \bar{d} d \rangle$.  Neglecting terms of order $m^2_{u,d}$,
identifying $f$ with the empirical pion decay constant $f_{\pi} = 92.4$ MeV, 
and inserting $m_u + m_d
\simeq 12$ MeV \cite{PP,Ioffe1}, 
one obtains
\begin{equation}
\langle \bar{q} q \rangle \simeq - (240 \, MeV)^3 \simeq -1.8 \, fm^{-3} .
\label{cond}
\end{equation}
This condensate presents a measure of the degree of spontaneous 
chiral symmetry breaking. The non-zero pion mass, on the
other hand, reflects the explicit symmetry breaking by the small quark masses,
with $m^2_{\pi} \sim m_q$. 
\begin{figure}
\begin{center}
\includegraphics[scale=0.65,angle=0]{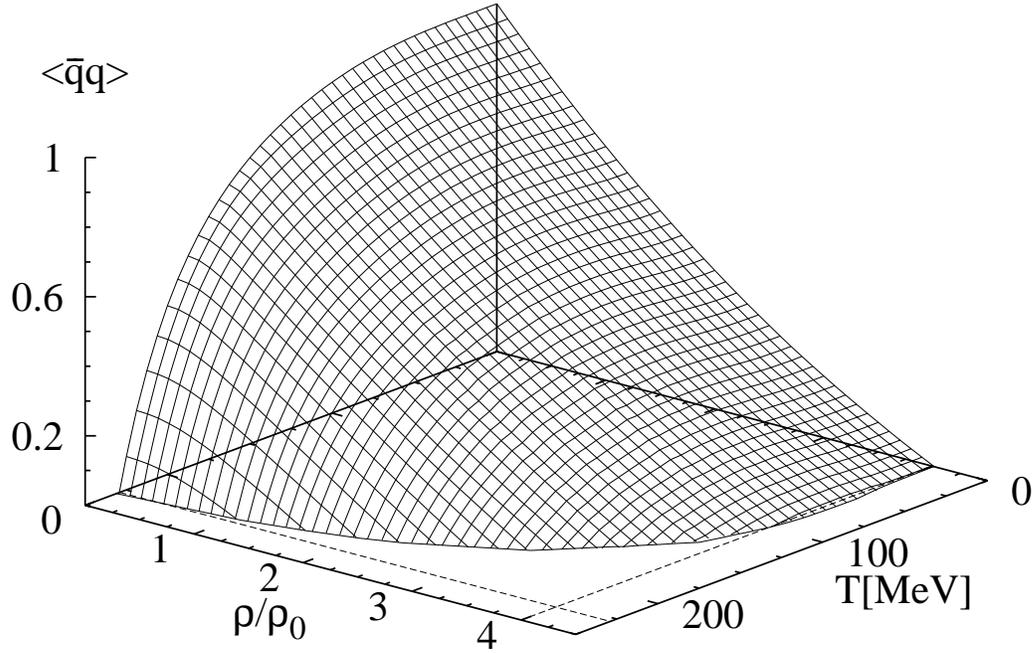}
\end{center}
\caption{Chiral condensate (in units of its vacuum value) as a function
of temperature and baryon density ($\rho_0 \simeq 0.16$ fm$^{-3}$ is the 
saturation density of nuclear matter).}
\label{fig:qqT}
\end{figure}

The strength of the scalar condensate Eq.~(\ref{cond}) is more than
an order of magnitude larger than the nuclear matter density at 
saturation: $\rho_0 \simeq 0.16$ fm$^{-3}$. Fig. \ref{fig:qqT} displays 
the temperature and baryon-density dependence of the chiral 
condensate in units of its vacuum value. At relatively low temperatures  
which are characteristic for low-energy nuclear physics, 
we notice an approximately linear dependence on baryon density 
up to $\rho\simeq\rho_0$, and even beyond. The strength of 
the quark condensate at normal nuclear densities is reduced by 
about one third from its vacuum value. The temperature dependence
is much weaker, at least up to $T\le 100$ MeV, far above the 
temperatures that can be sustained by ordinary nuclei.
The density-dependent changes of the condensate structure 
in the presence of baryonic matter are a source of strong scalar and vector 
fields experienced by the nucleons. At nuclear matter saturation density
several hundred MeV of scalar attraction are compensated by an almost equal 
amount of vector repulsion. The sum of the condensate nucleon fields
almost vanishes in infinite homogeneous nuclear matter.

\subsection{Chiral Effective Field Theory}
\label{sec:QCD-EFT}
Chiral EFT is a low-energy approximation to QCD. 
In the hadronic, low-energy phase, the active degrees of freedom of QCD are 
not elementary quarks and gluons, but rather mesons and baryons. The EFT
describes the active light particles as collective degrees of freedom,  
and the heavy particles are treated as static sources. The QCD Lagrangian  
is replaced by an effective Lagrangian which includes all relevant
symmetries of the underlying fundamental theory. Chiral EFT can then be used 
to calculate physical processes in terms of an expansion in $p/\Lambda$, where
$p$ denotes momenta or masses smaller than a certain momentum scale $\Lambda$.

In the meson sector (baryon number $B = 0$) the elementary quarks and 
gluons are replaced by Goldstone bosons. In the case of interest here, for 
two quark flavours ($N_f = 2$), the pseudoscalar Goldstone  
pions are represented by a unitary $2 \times 2$ matrix field $U (x) \in
SU(2)$ which collects the three isospin components $\pi_a (x)$ ($a=1,2,3$), 
and transforms under chiral rotations $\psi_R \to R\psi_R$ 
and $\psi_L \to L\psi_L$, as: 
\begin{equation}
U \to L U R^\dagger \quad {\rm with} \quad U (x) = \exp[i \tau_a \phi_a(x)] \; ,
\end{equation}
and $\phi_a = \pi_a/f$, where $f$ is the pion decay constant in the chiral limit.
Small corrections to $f$, of the order of the $u$- and $d$- quark masses, 
yield the physical pion decay constant $f_\pi = f + {\cal O} (m_{u,d}) = 92.4$ MeV.

Pions interact weakly at low energy: if $| \pi
\rangle = Q^A | 0 \rangle$ is a massless state with $H | \pi \rangle = 0$, then
a state $| \pi^n \rangle = (Q^A)^n | 0 \rangle$ with $n$ pions is
also massless since the axial charges $Q^A$ all commute with the full
Hamiltonian $H$. Interactions between pions must therefore vanish at
zero momentum and in the chiral limit.

It is easy to show that 
\begin{equation}
{\rm Tr} [ \partial^{\mu} U\partial_{\mu} U^{\dagger} ] 
\quad \rightarrow \quad 
{\rm Tr} [ L\partial^{\mu} U R^{\dagger} R \partial_{\mu} U^{\dagger} L^{\dagger}] 
= {\rm Tr} [ \partial^{\mu} U\partial_{\mu} U^{\dagger} ]
\end{equation}
is invariant under chiral transformations. 
The QCD Lagrangian of Eq.~(\ref{QCD}) is replaced by an effective Lagrangian 
which involves an expansion in the field $U (x)$ and its derivatives:
\begin{equation}
{\cal L}_{QCD} \to {\cal L}_{eff} (U, \partial U, \partial^2 U, ...) .
\label{Lagr}
\end{equation}
Momenta and derivatives are equivalent for this expansion. Pions do not interact 
unless they carry non-zero four-momentum, so the low-energy
expansion of Eq.~(\ref{Lagr}) is an ordering in powers of $\partial_{\mu} U$. 
Only even numbers of derivatives are allowed, because the Lagrangian has to 
be a Lorentz sacalar:
\begin{equation}
{\cal L}_{eff} = {\cal L}_\pi^{(0)} + {\cal L}_\pi^{(2)} + {\cal L}_\pi^{(4)} + ...
\end{equation}
However, since $U$ is unitary, ${\cal L}_\pi^{(0)}$ can only be a constant.
The expansion converges if the momenta are much smaller than a characteristic 
scale. In the case of chiral perturbation theory, 
this scale is $\Lambda_\chi = 4\pi f_\pi = 1161$ MeV, which provides a natural 
separation between ``light" and ``heavy" degrees of freedom.
The leading term, which is also know as the ``non-linear sigma
model", involves two derivatives:
\begin{equation}
{\cal L}_\pi^{(2)} = \frac{f^2}{4} Tr [ \partial_{\mu} U^{\dagger} \partial^{\mu} U] .
\end{equation}
A completely symmetric Lagrangian corresponds to massless Goldstone bosons, 
whereas the physical pions are massive. This is taken into account by adding a 
symmetry breaking term linear in the quark mass matrix $m$:
\begin{equation}
{\cal L}_\pi^{(2)} = \frac{f^2}{4} Tr [\partial_{\mu} U^{\dagger} \partial^{\mu} U]
+ \frac{f^2}{2} B_0 \, Tr [m (U + U^{\dagger})] \; .
\end{equation}
The symmetry breaking mass term is small and it can be treated
perturbatively, together with the power series in momentum. The index 
2 denotes either two derivatives or one quark mass term, and
this Lagrangian contains already all possible terms up to second order.
At fourth order, the terms permitted by symmetries are :
\begin{equation}
{\cal L}_\pi^{(4)} = \frac{l_1}{4} ( Tr [\partial_{\mu} U^{\dagger}
\partial^{\mu} U])^2 +
\frac{l_2}{4} Tr [\partial_{\mu} U^{\dagger} \partial_{\nu} U] Tr
[\partial^{\mu} U^{\dagger} \partial^{\nu} U] \; ,
\end{equation}
and the constants $l_1, l_2$ (following the standard notation of
ref.~\cite{GL}) must be determined by experiment. 
The fourth order Lagrangian ${\cal L}_\pi^{(4)}$ contains also symmetry
breaking terms, with additional constants
$l_i$ not determined by chiral symmetry.

An expansion of ${\cal L}^{(2)}$ to terms quadratic in the pion field 
yields the pion Lagrangian plus a constant vacuum contribution:
\begin{equation}
{\cal L}_\pi^{(2)} = (m_u + m_d) f^2 B_0 + \frac{1}{2} \partial_{\mu} \pi_a
\partial^{\mu} \pi_a - \frac{1}{2} (m_u + m_d) B_0 \pi^2_a + 0 (\pi^4)\; .
\label{L2}
\end{equation}
The first term in Eq.~(\ref{L2}) corresponds to the constant
shift of the vacuum energy density by the non-zero quark masses, 
and the pion mass term is identified as $m^2_{\pi} = (m_u + m_d)
B_0$. Using the GOR relation Eq.~(\ref{GOR}), one finds 
$- f^2 B_0 =  \langle \bar{q} q \rangle$.

In addition to the expansion of the effective Lagrangian in terms with an 
increasing number of derivatives and quark mass terms, Chiral Perturbation 
Theory (ChPT) is defined by a systematic method of estimating the importance 
of diagrams generated by the interaction terms -- the power counting scheme. 
Diagrams are classified according to the power of the generic variable Q 
(three-momentum or energy of the pion, or the pion mass $m_{\pi}$), and 
the small expansion parameter is $Q/\Lambda_\chi$.

In order to apply ChPT to the nucleon sector, pion-nucleon interactions must  
consistently be included in
the computational framework based on the low-energy 
expansion of the effective Lagrangian. 
The free-nucleon Lagrangian reads:
\begin{equation}
{\cal L}_0^N = \bar{\Psi}_N(i\gamma_{\mu}\partial^{\mu} - M_0)\Psi_N
\label{Nuc}
\end{equation}
where the Dirac spinor $\Psi_N(x) = (p,n)^T$ denotes
the isospin-$1/2$ doublet of proton and neutron, and  $M_0$
is the nucleon mass in the chiral limit.
The low-energy effective Lagrangian for pions 
interacting with a nucleon is obtained by replacing the 
pure meson Lagrangian by ${\cal L}_{eff} = {\cal L}_{\pi}^{(2)} + 
{\cal L}_{\pi}^{(4)} + ... + {\cal L}_{eff}^N $, which also
includes the nucleon field. The additional pion-nucleon term is 
expanded in powers of derivatives (momenta) and quark masses: 
\begin{equation}
{\cal L}_{eff}^N = {\cal L}_{\pi N}^{(1)} + {\cal L}_{\pi N}^{(2)} ~...
\end{equation}
and the structure of the $\pi N$
effective Lagrangian at each order is again determined by chiral symmetry.
The leading term, ${\cal L}_{\pi N}^{(1)}$, reads:
\begin{equation}
{\cal L}_{\pi N}^{(1)} =  \bar{\Psi}_N[i\gamma_{\mu}(\partial^{\mu} 
- iv^{\mu}) + g_A \gamma_{\mu}\gamma_5 a^{\mu} - M_0]\Psi_N \; .
\label{LPN}
\end{equation}
It includes the  
chiral covariant derivative with the vector coupling between 
the pions and the nucleon, and the axial-vector coupling. At this order 
the Lagrangian contains two parameters not determined by chiral symmetry:
the nucleon mass $M_0$ and the axial-vector coupling constant $g_A$.
The vector and axial vector fields can be expressed in terms of the chiral matrix
 $\xi \equiv \sqrt{U}$:
\begin{eqnarray}
v^{\mu} & = & \frac{i}{2}(\xi^{\dagger}\partial^{\mu}\xi +  
\xi\partial^{\mu}\xi^{\dagger}) = -\frac{1}{4f^2}\varepsilon_{abc}\tau_a~\pi_b
~\partial^{\mu}\pi_c + ...~~, \\
a^{\mu} & = & \frac{i}{2}(\xi^{\dagger}\partial^{\mu}\xi -  
\xi\partial^{\mu}\xi^{\dagger}) = -\frac{1}{2f^2}\tau_a~
\partial^{\mu}\pi_a + ...~~,
\end{eqnarray}
and expanded in powers of the pion field and its derivatives. 

At next-to-leading order, in ${\cal L}_{\pi N}^{(2)}$ the symmetry breaking 
quark mass term appears, and shifts the nucleon mass from
its value in the chiral limit to the physical mass:
\begin{equation}
M_N= M_0+\sigma_N \; .
\end{equation}
The ``sigma term" $\sigma_N$ 
represents the contribution from explicit chiral symmetry breaking to the nucleon mass 
\begin{equation}
\sigma_N=\! \sum_{q=u,d}  m_q \frac{{\rm{d}} M_N}{{\rm{d}}m_q} = \langle N |m_u\, \bar{u}u+m_d\,   \bar{d}d| N \rangle   \; ,
\label{sigma}
\end{equation}
through the non-vanishing $u$- and $d$-quark masses. Its empirical value 
$\sigma_N = (45 \pm 8)$ MeV,
has been deduced from low-energy pion-nucleon scattering data \cite{GLS}. 

The $\pi N$ effective Lagrangian, expanded to second order
in the pion field, reads:
\begin{eqnarray}
{\cal L}_{eff}^{N} & = & \bar{\Psi}_N(i\gamma_{\mu}\partial^{\mu} 
- M_N)\Psi_N - \frac{g_A}{2f_{\pi}} \bar{\Psi}_N\gamma_{\mu}\gamma_5 
\mbox{\boldmath $\tau$}\Psi_N\cdot\partial^{\mu}\mbox{\boldmath $\pi$}  \\
                   &   & -\frac{1}{4f_{\pi}^2}
 \bar{\Psi}_N\gamma_{\mu} 
\mbox{\boldmath $\tau$}\Psi_N\cdot\mbox{\boldmath $\pi$}\times
\partial^{\mu}\mbox{\boldmath $\pi$}
+ \frac{\sigma_N}{f_{\pi}^2} \bar{\Psi}_N\Psi_N\mbox{\boldmath $\pi$}^2 
+ ...~~,\nonumber
\end{eqnarray}
The microscopic origin of the empirical axial-vector coupling constant 
$g_A= 1.267 \pm 0.003$, determined from   
neutron beta decay $(n \rightarrow p e \bar{\nu})$,
is in the substructure of the nucleon, that cannot be resolved at the level 
of a low-energy effective theory. 

For a detailed review of baryon chiral perturbation theory we refer the 
reader to Ref.~\cite{BKM}. In applications of in-medium ChPT to the nuclear 
many-body problem (cf. Sec.~\ref{sec:NM}), 
the calculations will be based on the leading terms
in the effective Lagrangian, linear in the derivative
$\partial^{\mu} {\bm{\pi}}$ of the pion field.

\subsection{QCD Sum Rules}
\label{sec:QCD-SR}
The QCD sum rule approach \cite{SVZ.79} has been designed to interpolate 
between the perturbative (short-distance) and non-perturbative (large distances) 
energy sectors. It provides qualitative and quantitative information on hadron 
parameters (masses and coupling constants) from vacuum expectation values 
of QCD operators. 
 
The basic quantity is a momentum-space correlation function (correlator) of color-singlet
currents: 
\begin{equation}
\Pi(q^2) = i \int d^4 x~e^{iq\cdot x}~ \langle 0 | {\cal T} {\cal J} (x)  {\cal J}^\dagger 
(0) | 0 \rangle 
\end{equation}
where ${\cal J} (x)$ is a current composite operator of quark fields, 
$ {\cal T}$ denotes the time-ordered product, and $ | 0 \rangle$ is the physical 
non-perturbative vacuum. The singularities of the correlator correspond to 
physical excitations with the quantum numbers of the current ${\cal J}$. The 
sum rule analysis proceeds by introducing the {\it spectral function} of the 
correlator defined by the dispersion relation
\begin{equation}
\Pi(q^2) =  \int_0^\infty  ds ~{{\rho(s)}\over{s^2 - q^2 - i\epsilon} } \; ,
\label{rho}
\end{equation}
where 
\begin{equation}
\Theta(q_0) \rho(q^2) = {1\over \pi} {\rm Im} \Pi(q^2) =
(2\pi)^3 \sum_k ~\delta^4(p_k - q)~ | \langle 0 | {\cal J} (0)| k \rangle |^2
\label{rho2}
\end{equation}
describes the spectrum of physical intermediate states. For a fixed three-momentum 
${\bm q}$, the spectral function measures the intensity at which energy is 
absorbed from the current at different frequencies. 

The second step in the sum rule approach is a direct calculation of the correlator 
using the operator product expansion (OPE), which provides a QCD approximation 
to the correlator that is applicable at large spacelike $q^2$. The basic concept is 
the expansion of a time-ordered product of two local operators at short 
distances in terms of a complete set of local operators:
\begin{equation}
 {\cal T} A(x) B(0) ~\stackrel{x \to 0} {=}~\sum_n C_n^{AB} (x){\hat O}_n(0)  \; .
\end{equation}
The c-number coefficients $C_n^{AB}$ of the expansion are called 
Wilson coefficients. Therefore, in momentum-space:
\begin{equation}
\Pi(q^2) = i \int d^4 x~e^{iq\cdot x}~ \langle 0 | {\cal T} {\cal J} (x)  {\cal J}^\dagger 
(0) | 0 \rangle = \sum_n C_n (q^2)  \langle 0 |{\hat O}_n | 0 \rangle
\label{OPE}
\end{equation}
The operators ${\hat O}_n$ are built from quark and gluon fields. The operators of 
lowest dimension are $m_q {\bar q}q$ and $G_{\mu\nu}^a G_a^{\mu\nu}$. 
At the next order one finds 
four-quark operators and combinations of quark and gluon fields. The right-hand side 
of Eq.~(\ref{OPE}) is an expansion in vacuum expectation values (condensates) of 
the operators ${\hat O}_n$. The OPE is in inverse powers of $Q^2 \equiv - q^2$, and 
converges rapidly for large spacelike values of $Q^2 > 0$. 

The OPE separates short-distance and long-distance physics. In QCD this separation 
is between perturbative physics (the coefficients) and non-perturbative physics 
(the condensates). The Wilson coefficients decrease with inverse powers of 
$Q^2$ for each higher dimension operator, and can be calculated in QCD 
perturbation theory. The non-perturbative physics is contained in the condensates.

The fundamental relation of the QCD sum rule method is obtained by equating 
the dispersion relation for the correlator $\Pi(q^2)$ and its OPE representation, 
each evaluated at large spacelike $Q^2$:

\begin{eqnarray}
\Pi(q^2) &=&  {1\over \pi} \int_0^\infty  ds ~{{{\rm Im} \Pi(s)}\over{s^2 + Q^2 } }  +
{\rm polynomial} \nonumber \\
&=& C_0(-Q^2) + C_1(-Q^2) \langle m_q {\bar q}q \rangle + C_2(-Q^2)
 \langle G_{\mu\nu}^a G_a^{\mu\nu} \rangle + \cdots
 \label{SR}
\end{eqnarray}
In applications of QCD sum rules the convergence of both sides of Eq.~(\ref{SR}) is  
improved by applying the Borel transform -- a differential operation defined, for 
a given function $f(Q^2)$, by the relation:
\begin{equation}
{\cal B} f(Q^2) = \lim_{Q^2, n \to \infty}
{{(Q^2)^{n}}\over{(n-1)!}} \left ( - {{d}\over{dQ^2}} \right )^n f(Q^2) \;,
\end{equation}
with $Q^2/n \equiv {\cal M}^2$. The transform ${\cal B} f(Q^2) \equiv \hat f({\cal M}^2)$ 
depends on the parameter ${\cal M}$ -- the Borel mass. Any simple 
polynomial in $Q^2$ is eliminated by the Borel transform, 
because it cannot survive the infinite number of differentiations, 
and an inverse power of 
$Q^2$ is replaced by an exponentially decreasing function of the Borel mass:
\begin{equation}
{\cal B} {1\over {s+Q^2}} = {1\over {\cal M}^2} e^{-s/{\cal M}^2}\; .
\end{equation}
Therefore, the Borel transform applied to Eq.~(\ref{SR}) simultaneously eliminates the 
subtraction terms accompanying the dispersion relation and any divergent 
polynomial from the OPE. Higher-order terms in the OPE, which contain 
inverse powers of $Q^2$, are suppressed factorially by the Borel transform:
\begin{equation}
{\cal B} {1\over {(Q^2)^n}} = {1\over (n-1)! ({\cal M}^2)^n} \; .
\end{equation}

The correlator in spectral form can be evaluated by introducing a phenomenological 
model for the spectral function Eq.~(\ref{rho2}). Maching the Borel transforms of the 
OPE and phenomenological description of the correlator, yields the sum rules for
each invariant function. The leading order result for the nucleon mass in vacuum 
is the Ioffe's formula:
\begin{equation}
M_N = - {{8 \pi^2}\over{{\cal M}^2}}\langle {\bar q}q \rangle + \cdots
\label{Io}
\end{equation}
For ${\cal M} \approx 1$ GeV, one finds $M_N \approx 1$ GeV. Ioffe's
formula is not very accurate and the contribution from higher-dimensional
condensates should also be included. Nevertheless, it demonstrates that
the scale of the nucleon mass is largely determined by the scalar quark condensate, 
i.e. by the spontaneous chiral symmetry breaking mechanism. 

In section \ref{sec:NM-QCDSR} we will show how QCD sum rules at finite densities 
can be used to relate the 
leading changes of the scalar quark condensate $\langle \bar q q \rangle$ and 
of the quark density $\langle q^\dagger q \rangle$ at finite baryon density, 
with the scalar and vector self-energies of a nucleon in the nuclear medium.

\section{Chiral Dynamics and Nuclear Matter}
\label{sec:NM}
In this section we begin to approach one of the central problems of modern 
theoretical nuclear physics: how to establish a
relationship between low-energy, non-perturbative QCD and the rich nuclear phenomenology, which includes both nuclear matter and finite nuclei.  
We will first focus on infinite nuclear matter and analyze the equations of state of 
symmetric and asymmetric matter, as well as pure neutron matter, in the framework 
of chiral effective field theory. It will be shown that
nuclear binding and saturation arise primarily from chiral (pionic) 
fluctuations in combination with 
Pauli blocking effects and three-nucleon (3N) interactions, 
superimposed on the condensate background fields 
and calculated according to the rules of in-medium chiral perturbation 
theory (ChPT).

The key element of chiral EFT is a separation of long- and short- 
distance dynamics and an ordering scheme in powers of small momenta. In a medium, 
the relevant quantity for this expansion is the Fermi momentum $k_f$, which defines 
the nucleon density via the relation:
\begin{equation}
\rho(k_f) = 4 \int_0^{k_f} {{d^3p}\over {(2\pi)^3}} = {{2k_f^3}\over{3\pi^2}}
\end{equation} 
At nuclear matter saturation density ($\rho_0 \simeq 0.16$ fm$^{-3}$): 
$k_{f 0} \simeq 2m_{\pi}$, so the Fermi momentum and the pion mass are of 
comparable magnitude at the densities of interest. This means that pions should be included as {\it explicit} degrees of freedom in the description of nuclear 
many-body dynamics. The relevant observable is the 
energy per particle $\bar E(k_f)$, i.e. the ratio of energy density and 
particle density, with the free nucleon mass $M$ subtracted.

Both $k_f$ and $m_{\pi}$ are small compared to the characteristic
chiral scale, $4 \pi f_{\pi} \simeq 1.2$ GeV. 
Consequently, the equation of state (EOS) of
nuclear matter as given by ChPT will be represented as an
expansion in powers of the Fermi momentum. The expansion coefficients are
non-trivial functions of $k_f/ m_{\pi}$, the dimensionless ratio of the two
relevant scales inherent to the problem.
\begin{figure}
\begin{center}
\includegraphics[scale=1.25,angle=0]{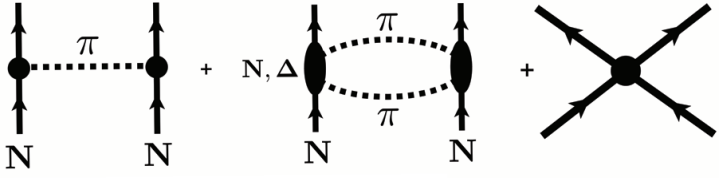}
\end{center}
\caption{NN amplitude in chiral effective field theory:  one-pion exchange, two-pion exchange (including $\Delta$ isobar intermediate states), and contact terms 
representing short-distance dynamics.}
\label{fig:PI-EX}
\end{figure}

\subsection{In-medium Chiral Perturbation Theory}
\label{sec:NM-ChPT}
In the chiral EFT framework nuclear dynamics is described by a low-energy 
effective representation of QCD. The relevant degrees of freedom are 
pions and nucleons, and observables are calculated in chiral perturbation 
theory, as described in Sec.~\ref{sec:QCD-EFT}. The chiral effective Lagrangian 
Eq.~(\ref{LPN}),  generates the basic pion-nucleon coupling terms.
In the nuclear medium, however, all nucleon states below some Fermi momentum 
$k_f$ are occupied. Instead of the ``empty" vacuum $|0\rangle$, the ground 
state is the filled Fermi sea $|\phi_0 \rangle$, and the nucleon propagator 
in coordinate space changes to:
\begin{equation}
S_F^0(x-y) = \langle 0 | {\cal T}\psi (x) {\bar \psi}(y) |0\rangle \quad 
\to \quad S_F(x-y) = \langle \phi_0 | {\cal T}\psi (x) {\bar \psi}(y) |\phi_0\rangle \; .
\end{equation}
In momentum space there is a very useful additive decomposition of the in-medium nucleon propagator. For a relativistic nucleon with
four-momentum $p_{\mu} = (p_0, \vec{p} \,)$ it reads
\begin{equation}
(\not \!p + M) \left\{ \frac{i}{p^2 - M^2 + i \varepsilon} - 2 \pi \delta (p^2
- M^2) \theta (p_0) \theta(k_f - | \vec{p}\, |) \right\}\; .
\label{in-medium}
\end{equation}
The second term is the medium insertion which accounts for the fact that the
ground state of the system has changed from an ``empty" vacuum to a filled Fermi
sea of nucleons. Diagrams can then be organized systematically  in the number
of medium insertions, and an expansion is performed in leading inverse powers
of the nucleon mass, consistent with the $k_f$-expansion.

The NN amplitude is represented by the diagrams shown in 
Fig.~\ref{fig:PI-EX}. One- and two-pion exchange processes 
are treated explicitly. They govern the long-range interactions at 
distance scales $d > 1/k_f$ fm, whereas short-range dynamics
is not resolved at nuclear Fermi momentum 
scales, and can be subsumed in contact interaction terms.

We notice that the two-pion exchange diagram in Fig.~\ref{fig:PI-EX} includes 
not only nucleon intermediate states, but also single and double virtual 
$\Delta(1232)$-isobar excitations. In fact, in addition
to $k_f$ and $m_\pi$, a third relevant ``small" scale is the mass difference 
$\delta M = M_\Delta - M_N \simeq 0.3$ GeV between the $\Delta(1232)$ and the nucleon. The strong spin-isospin transition from the nucleon to the $\Delta$ isobar 
must therefore be included as an additional ingredient in the in-medium ChPT 
calculation, so that the expansion coefficients of the nuclear matter equation of state
become functions of both $k_f / m_\pi$ and $m_\pi / \delta M$.

Initial applications of in-medium ChPT to nuclear dynamics \cite{LFA.00,Kai.01jx} 
have shown that basic properties of the equation of state for isospin-symmetric nuclear matter can already be reproduced by $\pi N$-dynamics alone, without introducing 
virtual $\Delta$-excitations. In the next section, therefore, we will first discuss nuclear 
matter properties calculated in a ``one-parameter" chiral approach, which considers
only $\pi N$-dynamics. In Sec. \ref{sec:NM-Delta} we will show that the explicit inclusion 
of $\Delta(1232)$ degrees of freedom leads to a systematic improvement of 
the chiral EFT description of nuclear matter, especially in the isovector channel. 
\subsection{Chiral ``one-parameter" Nuclear Matter Equation of State}
\label{sec:NM-EOS}

\begin{figure}
\begin{center}
\hspace{2.5mm}
\includegraphics[scale=0.6,angle=0]{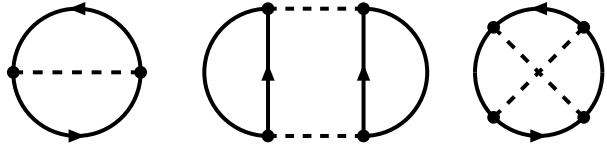}
\end{center}
\caption{One-pion exchange Fock diagram, and iterated one-pion 
exchange Hartree and 
Fock diagrams.}
\label{fig:pi-nm-a}
\end{figure}

\begin{figure}
\begin{center}
\hspace{2.5mm}
\includegraphics[scale=0.65,angle=0]{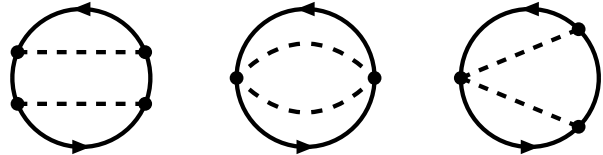}
\end{center}
\caption{Irreducible two-pion exchange diagrams.}
\label{fig:pi-nm-b}
\end{figure}

The in-medium ChPT three-loop calculation of the
energy per particle of symmetric nuclear matter has been carried out 
in Ref.~\cite{Kai.01jx}, including contributions up to order ${\cal O}(k_f^5)$.
These contributions are: the kinetic energy (contributing at order ${\cal O}(k_f^2)$),  chiral one-pion exchange (${\cal O}(k_f^3)$), iterated one-pion exchange 
(${\cal O}(k_f^4)$),  and irreducible two-pion exchange (${\cal O}(k_f^5)$). 
The corresponding closed-loop diagrams are shown in Fig.~\ref{fig:pi-nm-a} 
(the one-pion exchange Fock diagram, and iterated one-pion 
exchange Hartree and Fock diagrams),  
and in Fig.~\ref{fig:pi-nm-b} (irreducible two-pion exchange diagrams).
The $2\pi$ Fock diagram has both a reducible part contributing to the 
iterated $1\pi$ exchange, and an irreducible two-pion exchange part \cite{KBW.97}. 

The calculation involves one single momentum-space cut-off
$\Lambda$ which encodes dynamics at short distances not resolved explicitly in
the effective low-energy theory. This high-momentum scale $\Lambda$ is the only
free parameter. Note that instead of a momentum-space cut-off,  
one could equivalently use dimensional regularization, remove divergent loop 
integrals, and replace them by adjustable $NN$ contact terms which then 
parameterize the unresolved short-distance physics.

The saturation of nuclear matter can simply be illustrated in the 
exact chiral limit: $m_\pi =0$. The saturation
mechanism can already be demonstrated by truncating the one- and 
two-pion exchange diagrams at order ${\cal O}(k^4_f)$.
The energy per particle, $\bar{E}(k_f) = E(k_f)/A$, of isospin symmetric nuclear 
matter is given in powers of the Fermi momentum $k_f$, and can simply be
parameterized: 
\begin{equation}
\bar{E} (k_f) = \frac{3 k^2_f}{10 \, M_N} - \alpha \frac{k^3_f}{M_N^2} + \beta
\frac{k^4_f}{M_N^3} \; .
\label{simple_eos}
\end{equation}
The first term is the kinetic energy, and the result for $\alpha$ in the chiral limit is:
\begin{equation}
\alpha = \left( \frac{g_{\pi N}}{4 \pi} \right)^2
\left[\frac{10 \Lambda}{M_N} \left( \frac{g_{\pi N}}{4 \pi} \right)^2
 - 1\right] \; ,
\end{equation}
where $M_N$ is the nucleon mass, and $g_{\pi N} = g_A M_N /f_{\pi}$ is 
the empirical $\pi N$ coupling constant: $g_{\pi N} = 13.2$.
The strongly attractive leading $k^3_f$-term is accompanied 
by the weakly repulsive one-pion exchange Fock term. 
The $k^3_f$-contribution to $\bar{E} (k_f)$ would lead to collapse, 
were it not for the stabilizing $k^4_f$-term controlled by the coefficient:
\begin{equation}
\beta = \frac{3}{70} \left( \frac{g_{\pi N}}{4 \pi} \right)^4 ( 4 \pi^2 + 237 -
24 \ln 2) - \frac{3}{56} = 13.55 \; .
\label{beta}
\end{equation}
The two-pion exchange between nucleons generates the attraction proportional 
to $k_f^3$ in the energy per particle. The Pauli blocking of intermediate nucleon
states in these $2\pi$ exchange processes produces the repulsive $k_f^4$ term. 
The parameter-free ChPT value $\beta=13.55$ in the chiral limit, is very 
close to the value ($\simeq 12.2$) that one obtains by
adjusting the simple equation of state Eq.~(\ref{simple_eos}) to empirical 
saturation properties (binding energy and saturation density) \cite{Kai.01jx}.
For values of the short-distance scale $\Lambda$ between $0.5$ and 
$0.6$ GeV the EOS has a stable minimum of $\bar{E} (k_f)$ in the 
empirical range of density and binding energy.

\begin{figure}
\begin{center}
\vspace{0.5 cm}
\includegraphics[scale=0.575,angle=0]{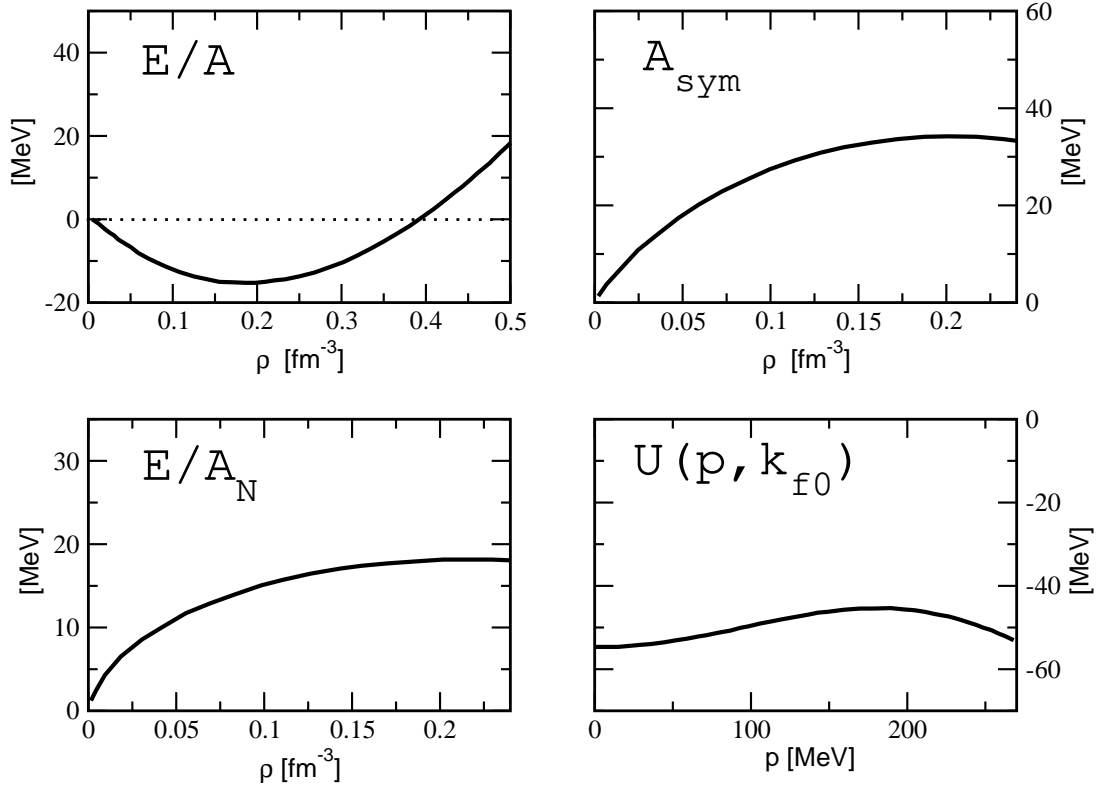}
\end{center}
\caption{Energy per particle of symmetric nuclear matter as function of 
the nucleon density, determined by chiral one- and two-pion exchange 
up to order  ${\cal O}(k_f^5)$ (upper left panel). The corresponding 
asymmetry energy as function of the nucleon density (upper right), 
and the density dependence of the energy per particle of pure 
neutron matter (lower left). The momentum dependence of the 
real part of the single-nucleon potential at nuclear saturation 
Fermi momentum $k_{f0} = 272.7$ MeV, is shown 
in the lower right panel.}
\label{fig:EOS}
\end{figure}

The full 3-loop chiral dynamics result for $\bar{E} (\rho)$ in symmetric
nuclear matter with nucleon density $\rho(k_f) =  {{2k_f^3}/({3\pi^2}})$,
using $m_{\pi} = 135$ MeV, and including all calculated terms up to 
order ${\cal O}(k_f^5)$, is shown in the upper left panel 
of Fig.~\ref{fig:EOS}.  The single-parameter of the calculation -- the 
cut-off scale $\Lambda = 0.65$ GeV, has been 
adjusted to the value $\bar{E}_0 = -15.3$ MeV at equilibrium,
and the resulting energy per particle $\bar E(\rho)$ 
has a minimum with this value of $\bar E_0$ at the density $\rho_0 = 0.178$ 
fm$^{-3}$ (corresponding to a Fermi momentum of $k_{f0}= 272.7\; {\rm MeV}
=1.382\,$fm$^{-1}$). The predicted compression modulus:
\begin{equation}
 K= k_{f0}^2\, {\partial^2 \bar E(k_f) \over \partial k_f^2}
\bigg|_{k_f=k_{f0}} = 255\,{\rm MeV} \; ,
\end{equation} 
is in quantitative agreement with the ``empirical" value 
$250\pm25$ MeV \cite{Bla.80,Vre.03}. It is also interesting to 
analyze the various contributions to the 
equilibrium value $\bar E_0=-15.26$ MeV \cite{Kai.01jx}.
Its decomposition into contributions from 
the kinetic energy and the three classes of diagrams 
($1\pi$-exchange Fock, iterated $1\pi$-exchange, and 
irreducible $2\pi$-exchange) reads: $\bar E_0=(23.40+18.24
-68.35+11.45)$ MeV. The nuclear matter binding energy 
can also be decomposed into contributions ordered in chiral powers 
${\cal O}(k_f^\nu), (\nu=2,3,4,5)$:
$\bar E_0 = (23.75 -154.54+124.61-9.08)$ MeV.

The simplest way to extend the in-medium ChPT calculation
to isospin asymmetric matter is to use the same cut-off for all 
isospin channels. The following substitution:
\begin{equation} \theta(k_f-|\vec p\,|) \quad \to \quad {1+\tau_3 \over 2}\,  
\theta(k_p-|\vec p\,|) +{1-\tau_3 \over 2}\, \theta(k_n-|\vec p\,|)\,,
\end{equation}
in the in-medium nucleon propagator Eq.~(\ref{in-medium}), takes into account 
the fact that the Fermi seas of protons and neutrons are no longer 
equally filled. $k_p$ and $k_n$ denote the 
Fermi momenta of protons and neutrons, respectively. Choosing $k_{p,n} = 
k_f(1\mp \delta)^{1/3}$ (with $\delta$ a small parameter) the nucleon density 
$\rho=\rho_p+\rho_n= (k_p^3+k_n^3)/3\pi^2= 2k_f^3/3\pi^2$ remains constant. The 
expansion of the energy per particle of isospin asymmetric  nuclear matter:
\begin{equation} 
\bar E_{as}(k_p,k_n)= \bar E(k_f) + \delta^2\, A(k_f)+ \cdots  \; ,
\label{asym}
\end{equation}
around the symmetry line ($k_p=k_n$, or $\delta=0$) defines the asymmetry energy
$A(k_f)$. Note that the parameter $\delta$ is equal to $(\rho_n-\rho_p)/
(\rho_n+\rho_p)$, or $(N-Z)/(N+Z)$. In the upper right panel of
Fig.~\ref{fig:EOS} we display the density dependence of $A(k_f)$ as determined 
by $\pi N$-dynamics up to ${\cal O}(k_f^5)$ and three-loop order. At the saturation 
point $k_{f0}= 272.7$ MeV: $A_0 \equiv A(k_{f0}) = 33.8$ MeV, in very good agreement 
with the empirical value: $A_0 = (30\pm 4)$ MeV. 
Extrapolation to higher density works roughly up to $\rho \simeq 1.5 \,
\rho_0$. At still higher densities one observes an unphysical downward bending 
of the asymmetry energy curve, indicating the limit of validity of the chiral expansion 
scheme restricted to pion-nucleon dynamics only, and with only a single 
momentum cut-off $\Lambda = 0.65$ GeV parameterizing the unresolved 
short-range dynamics.

The extreme of asymmetric nuclear matter is pure neutron matter. This system is 
unbound and its energy per particle increases monotonically with neutron density.  
To calculate the energy per particle, $\bar E_n(k_n)$ of neutron 
matter in the framework  of in-medium ChPT, it is sufficient to make the substitution 
\begin{equation} 
\theta(k_f-|\vec p\,|) \quad \to \quad {1-\tau_3 \over 2}\, 
\theta(k_n-|\vec p\,|)\; ,
\end{equation}
in the nucleon propagator Eq.~(\ref{in-medium}). Here $k_n$ denotes the
Fermi momentum of the neutrons, related to the neutron density 
by $\rho_n=k_n^3/3\pi^2$. The resulting EOS
of neutron matter is shown in the lower left panel of Fig.~\ref{fig:EOS}. 
The convex shape of the curve is generic and does not change much with
the cut-off $\Lambda$. When compared with sophisticated many-body 
calculations of neutron matter EOS,  one finds a rough 
agreement up to neutron densities of about $\rho_n=0.25\,$fm$^{-3}$. At higher
densities the chiral $\pi N$ neutron equation of state becomes unrealistic
because of the downward bending, inherited from
the asymmetry energy $A(k_f)$.

The in-medium three-loop calculation of the energy per particle defines the
(momentum dependent) self-energy of a single nucleon in nuclear matter up to
two-loop order. The momentum- and density-dependent (complex-valued) 
single-particle potential of nucleons in isospin symmetric nuclear matter 
has been calculated in Ref.~\cite{Kai.01ra}. The momentum dependence of the 
real part $U(p,k_{f0})$ of the single-particle potential evaluated up to order 
${\cal O}(k_f^5)$, at nuclear saturation Fermi momentum 
$k_{f0} = 272.7$ MeV, is shown 
in the lower right panel of Fig.~\ref{fig:EOS}. For a nucleon at rest, one finds 
$U(0,k_{f0}) = -53.2$ MeV, in agreement with the empirical optical-model potential 
$U_0 \approx -52$ MeV, deduced by extrapolation from elastic nucleon-nucleon 
scattering data. The momentum dependence of $U(p,k_{f0})$ translates into an
effective nucleon mass $M_N^\star (p)$:
\begin{equation}
{1\over {M_N^\star (p)}} = {1\over p} {\partial \over {\partial p}} 
\left [ T_{kin}(p) + U(p,k_{f0}) \right ] \; .
\end{equation}
The strong momentum dependence with the up- and down-bending
shown in Fig.~\ref{fig:EOS}, produces a 
negative slope of $U(p,k_{f0})$ at $p=k_{f0}$, resulting in a far too large 
nucleon mass at the Fermi surface. As we will show in the next section, 
a significant improvement of the momentum dependence of $U(p,k_{f0})$ 
is obtained by explicitly including diagrams involving virtual $\Delta$-excitations.

The realistic values for several nuclear matter properties 
(binding energy and saturation density, compressibility, asymmetry energy, 
depth of the single-nucleon potential), have been 
obtained with only one adjustable scale parameter -- 
the momentum cut-off $\Lambda$. These results 
demonstrate that the nuclear binding and saturation mechanism 
based on the combination of chiral one- and two-pion exchange, with 
short-distance dynamics encoded in a single high-momentum scale 
$\Lambda$,  presents a very good starting point for a description of the 
nuclear many-body problem based on chiral dynamics, at least at low 
nucleon densities $\rho \leq 0.2 $ fm$^{-3}$.
%
\subsection{The Role of Virtual ${\bm \Delta}$(1232) Excitations}
\label{sec:NM-Delta}
%
The chiral approach to nuclear matter 
has been extended and improved in Ref.~\cite{Fri.04}, by systematically 
including contributions from two-pion exchange 
with single and double virtual $\Delta(1232)$-isobar excitations. 
The spin-3/2 isospin-3/2 $\Delta(1232)$-resonance is the most prominent
feature of low-energy $\pi N$-scattering. Two-pion exchange 
between nucleons with excitation of virtual $\Delta$-isobars generates 
most of the empirical strong isoscalar central NN-attraction \cite{gerst.98}, 
which in phenomenological 
one-boson exchange models is often simulated by a fictitious  scalar 
''$\sigma$-meson'' exchange. In addition, as we have already emphasized
in Sec.~\ref{sec:NM-ChPT}, the delta-nucleon mass splitting 
$\delta M = 293$ MeV is of the same size as the Fermi momentum $k_{f0} \simeq 
2m_\pi$ at nuclear matter saturation density, and therefore pions and 
$\Delta$-isobars should both be treated as explicit degrees of freedom.

In Ref.~\cite{Fri.04} a calculation of isospin-symmetric and isospin-asymmetric 
nuclear matter EOS has been carried out, including
all effects from $2\pi$-exchange with virtual 
$\Delta$-excitation up to three-loop order in the energy density. 
The leading contributions from $2\pi$-exchange with virtual $\Delta$-excitation 
to the energy per particle are generically of fifth power in the small
momenta  ($k_f,m_\pi,\delta M$). 
The effects from irreducible $2\pi$-exchange (cf. Sec.~\ref{sec:NM-EOS})
are of the same order. However, since the
$\pi N\Delta$-coupling constant is about twice as large as the $\pi
NN$-coupling constant ($g_{\pi N \Delta} = 3g_{\pi N}/\sqrt{2}$),  
the $\Delta$-driven $2\pi$-exchange
effects should dominate. The contributions to the energy per particle can be 
classified as two-body terms and three-body terms, with the latter 
interpreted as Pauli-blocking effects on the two-body terms imposed by the 
filled Fermi-sea of nucleons.
\begin{figure}
\begin{center}
\includegraphics[scale=0.8,clip]{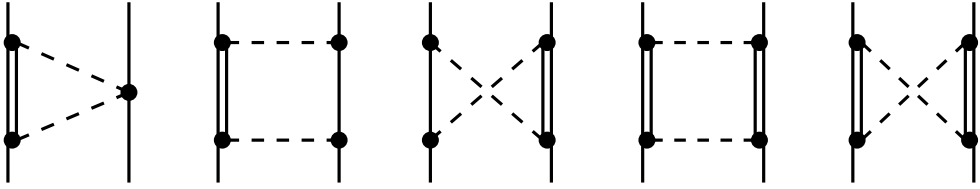}
\end{center}\vspace{-0.4cm}
\caption{One-loop two-pion exchange diagrams with single and double
$\Delta(1232)$-isobar excitations.}
\label{deltagraphs}
\end{figure}

Fig.~\ref{deltagraphs} displays the relevant one-loop triangle, box and 
crossed box two-pion exchange diagrams with single and double
$\Delta(1232)$-isobar excitations. By closing the two open nucleon lines to 
either two rings or one ring one gets (in diagrammatic representation) the 
Hartree or Fock contribution to the energy density of nuclear matter.
The NN potential in momentum space involves pion-loop diagrams which 
are in general ultra-violet divergent and require regularization (and renormalization). 
This procedure results in a few subtraction constants which encode unresolved 
short-distance NN-dynamics. The associated $k^3_f$ and $k^5_f$-terms in the 
energy per particle 
\begin{equation} 
\bar E^{(NN)}(k_f)= B_3 {k_f^3 \over M_N^2} + B_5 {k_f^5 \over
M_N^4}\; ,
\label{B3B5}
\end{equation}
are then adjusted to some empirical property of nuclear matter 
(e.g. the binding energy of $\approx 16$ MeV). $B_3$ and $B_5$ are chosen 
dimensionless,  and $M_N=939$ MeV is the nucleon mass.
Eq.~(\ref{B3B5}) replaces the cut-off regularization used 
in Sec.~\ref{sec:NM-EOS} and, in addition, also takes into account the effects 
from $p^2$-dependent contact terms. 

\begin{figure}
\begin{center}
\vspace{0.5 cm}
\includegraphics[scale=0.575,angle=0]{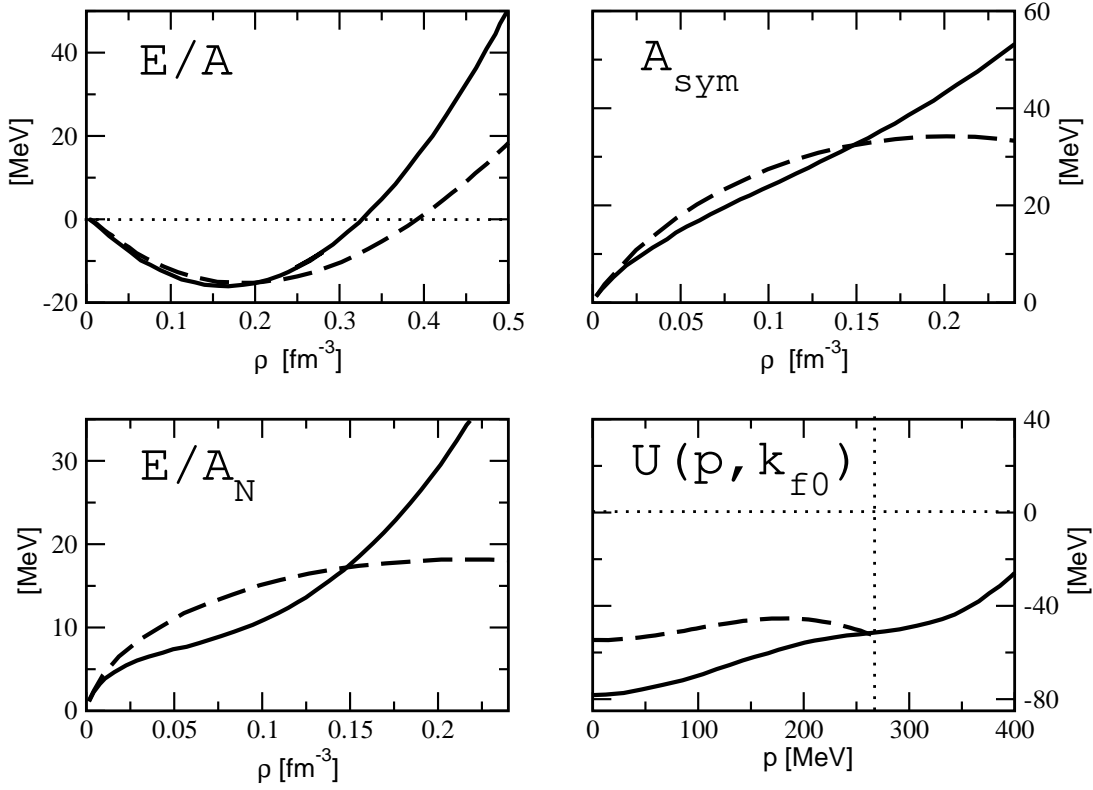}
\end{center}
\caption{Energy per particle of symmetric nuclear matter (upper left), and the  
asymmetry energy (upper right) as functions of 
the nucleon density. Energy per particle of pure 
neutron matter (lower left) as a function of neutron density, and the momentum 
dependence of the real part of the single-nucleon potential at nuclear saturation
(lower right). The dashed curves refer to the results described in 
Sec.~\ref{sec:NM-EOS}, with only pions and nucleons as active degrees of freedom.
The solid curves include the contributions from two-pion exchange 
with single and double virtual $\Delta(1232)$-isobar excitations. }
\label{fig:EOS-2}
\end{figure}

The resulting EOS for symmetric nuclear matter displays 
binding and saturation in a wide interval of values of 
the two adjustable parameters $B_3$ and $B_5$. However, a strong 
repulsive $\rho^2$-term from the three-body Hartree diagram 
leads to a too low saturation density $\rho_0$ and a too high 
nuclear matter compressibility: $K > 350$ MeV. This problem 
has been solved in a minimal way by introducing an attractive 
three-body contact term, 
\begin{equation} 
\bar E^{(NNN)}(k_f)= B_6 {k_f^6 \over M_N^5} \; .
\label{B6}
\end{equation}
The three parameters: $B_3$ and $B_5$ of the two-body contact term 
Eq.~(\ref{B3B5}), and $B_6$ of the three-body term Eq.~(\ref{B6}), 
determine not only the isospin-symmetric energy per particle $\bar E(k_f)$ 
at three-loop order, but also the the momentum dependence of the 
single-particle potential $U(p, k_f)$. With the following choice of the 
parameters: $B_3=-7.99$, $B_5 =0$, and $B_6 = -31.3$ \cite{Fri.04},
the minimum of the saturation curve $\bar E(k_f)$ is fixed to the
value $\bar E_0= -16.0$ MeV, the predicted value of the saturation 
density is $\rho_0= 0.157$ fm$^{-3}$,
corresponding to a Fermi momentum of $k_{f0} =261.6$ MeV 
$=1.326$ fm$^{-1}$, and the nuclear matter compressibility equals $K=304$ MeV, 
a somewhat high but still acceptable value. The 
density dependence of the energy per particle is shown in the 
upper left panel of Fig.~\ref{fig:EOS-2}, in comparison with the 
previous EOS calculated without the inclusion of virtual 
$\Delta$-excitations (cf. Sec.~\ref{sec:NM-EOS}). 
In that case the saturation density $\rho_0= 0.178\,$fm$^{-3}$ was 
somewhat too high, but the compressibility $K=255\,$MeV had a 
better value. The more pronounced increase of the solid
curve with the density $\rho$ results from the inclusion of higher order 
terms in the $k_f$-expansion.  

The solid curve in the lower right panel in Fig.~\ref{fig:EOS-2} represents 
the real part of the single-particle potential 
$U(p,k_{f0})$ at saturation Fermi momentum $k_{f0}= 261.6$ 
MeV, as a function of the nucleon momentum $p$, whereas the 
dashed curve shows the single-particle potential calculated with 
the restriction to only nucleon intermediate states in $2\pi$-exchange diagrams. 
We observe that with the chiral $\pi N\Delta$-dynamics
included, the real single-particle potential $U(p,k_{f0})$ increases
monotonically with the nucleon momentum $p$. The unphysical 
downward bending above $p=180$ MeV displayed by the dashed curve, 
is no longer present. The slope
at the Fermi surface $p=k_{f0}$ translates into an effective nucleon mass of
$M^*(k_{f0}) = 0.88 M$, a realistic value compared to $M^*(k_{f0}) \simeq 3M$ 
obtained in the previous calculation. The potential is now somewhat 
deeper: $U(0,k_{f0})= -78.2 $MeV, compared to $U(0,k_{f0}) = -53.2$ MeV 
for the case without virtual $\Delta$-excitations, but still realistic.

The equations of state of pure neutron matter are compared in the lower left 
panel of Fig. ~\ref{fig:EOS-2}. With the inclusion of chiral $\pi N\Delta$-dynamics, 
the short-range contribution reads:
\begin{equation} 
\bar E_n^{(NN)}(k_n)= B_{n,3} {k_n^3 \over M_N^2} + 
B_{n,5} {k_n^5 \over M_N^4}\; ,
\label{BN3BN5}
\end{equation}
where $B_{n,3}$ and $B_{n,5}$ are two new subtraction constants. 
Note that the Pauli-exclusion principle forbids a three-neutron
contact-interaction. The (dashed) solid curve represents the 
energy per particle $\bar E_n(k_n)$ of pure neutron matter 
as a function of the neutron density $\rho_n = k_n^3/3\pi^2$, 
calculated (without) with the inclusion of chiral $\pi N\Delta$-dynamics.
The short-range parameters $B_{n,3} = -0.95$ and $B_{n,5}=-3.58$ 
have been adjusted to the value of the asymmetry energy at saturation 
density $A(k_{f0})= 34 $ MeV (cf. upper right panel). The important 
result here is that the unphysical downward
bending of $\bar E_n(k_n)$ (displayed by the dashed curve)
disappears after the inclusion of $\pi N\Delta$ diagrams.
It has also been noted that up to
$\rho_n \approx 0.16$ fm$^{-3}$, the chiral $\pi N\Delta$ result for 
$\bar E_n(k_n)$ is similar to the standard EOS obtained with 
sophisticated many-body calculations based on free-space NN-interactions 
plus adjustable 3N-forces, whereas it becomes stiffer at
higher densities. Of course one should not expect the 
present model to work at Fermi momenta larger than 
$k_n \approx 350$ MeV. 

The improved isospin channel of the EOS is also manifest in the density 
dependence of the asymmetry energy $A(k_f)$ (upper right panel of 
Fig.~\ref{fig:EOS-2}), defined by Eq.~(\ref{asym}). The important feature 
is that the inclusion of chiral $\pi N\Delta$-dynamics (solid curve)
eliminates the unrealistic downward bending of the asymmetry $A(k_f)$ 
at higher densities $\rho>0.2$ fm$^{-3}$ (dashed curve). The value at 
saturation density $\rho_0=0.157$ fm$^{-3}$: $A(k_{f0})= 34.0$ MeV,
is consistent with the empirical values of the asymmetry energy: 
$(30 \pm 4)$ MeV.

In these last two sections we have shown that a microscopic approach to 
the nuclear many-body problem, based on chiral EFT, and which includes 
pions and $\Delta$-isobars as explicit degrees of freedom in in-medium 
ChPT calculations, successfully describes the equations of state of 
isospin-symmetric and isospin-asymmetric nuclear matter. 
In fact, a nuclear equation of state with realistic saturation properties 
(binding energy, saturation density, compression modulus)
can already be generated by the simplest version of 
the chiral framework, including only the 
contributions of one- and iterated one-pion exchange up to 
order ${\cal O}(k_f^4)$ and parameterizing all unresolved short-range 
dynamics with a simple momentum cut-off. In the calculation restricted 
to only nucleon intermediate states, i.e. without virtual $\Delta$-excitations, 
the asymmetry energy 
at the saturation point is well reproduced and neutron matter is predicted to be 
unbound, but the curves for both the asymmetry energy and the neutron matter
equation of state display an unphysical downward bending at higher densities.
The real part of the single-nucleon potential has the correct depth at momentum 
$p=0$, but the strong momentum dependence results in a far too large 
nucleon mass at the Fermi surface. These deficiencies of the in-medium ChPT 
calculation are cured by the explicit inclusion of contributions from two-pion 
exchange with single and double virtual $\Delta(1232)$-isobar excitations.
With two parameters per isospin channel for the unresolved short-range physics, the expansion is carried out up to three-loop order and up to fifth order in small scales
($k_f,m_\pi,\delta M$). The chiral $\pi N\Delta$-dynamics considerably improves 
both the momentum dependence of the single particle potential (realistic value 
for the effective mass), and the isovector properties of nuclear matter (density 
dependence of the asymmetry energy and neutron matter EOS).
\subsection{In-medium QCD Sum Rules}
\label{sec:NM-QCDSR}

Initial applications of the microscopic approach based on chiral 
$\pi N$ dynamics (not yet including the $\Delta(1232)$) 
to bulk and single-nucleon properties of finite nuclei  \cite{Fi.02,Fi.03},
have shown that chiral (two-pion exchange) fluctuations play
a prominent role in nuclear binding, but do not produce the strong
spin-orbit nuclear force responsible for the large energy spacings 
between spin-orbit partner single-nucleon states. The large 
effective spin-orbit potential in finite nuclei is generated by  
additional strong isoscalar scalar and vector nucleon self-energies,
induced by changes of the quark condensate and the quark density 
in the presence of baryonic matter.
The magnitude of these condensate background nucleon self-energies
can be estimated using leading-order in-medium QCD sum rules.

Applications of QCD sum rules at finite baryon density $\rho$ proceed 
analogously to the vacuum case (cf. Sec.~\ref{sec:QCD-SR}), except that now operators
such as $q^{\dagger}q$ also have non-vanishing ground state expectation values:
$\langle q^{\dagger}q \rangle_{\rho} =
3\rho/2$. In addition, the vacuum condensates vary with increasing density.
Thus in-medium QCD sum rules relate the changes of the 
scalar quark condensate and the quark density at 
finite baryon density, with the isoscalar scalar and vector self-energies 
of a nucleon in the nuclear medium. In leading order which should be valid at 
densities below and around saturated nuclear matter, the condensate 
part, $\Sigma_S^{(0)}$, of the scalar self-energy is expressed in terms of the density dependent chiral condensate as follows~\cite{CFG.91,DL.90,Jin.94}: 
\begin{equation} 
\Sigma^{(0)}_S = - \frac{8 \pi^2}{{\cal M}^2} [ 
\langle \bar{q} q \rangle_\rho - \langle \bar{q} q \rangle_0 ] 
 = - \frac{8 \pi^2}{{\cal M}^2}~\frac{\sigma_N}{m_u +m_d} \rho_s \; ,
\end{equation} 
where $\rho_s = \langle \bar{\Psi}\Psi \rangle$ is the nucleon scalar density. 
The chiral vacuum condensate $\langle \bar{q} q \rangle_0$ as a measure
of spontaneous chiral symmetry breaking in QCD is given in Eq.~(\ref{cond}). The difference between the vacuum condensate $\langle \bar{q} q \rangle_0$ 
and the one at finite density involves the nucleon sigma term, 
$\sigma_N \sim \langle N| m_q \bar{q} q |N \rangle$, to this order. 
The Borel mass scale ${\cal M} \approx 1$GeV roughly separates 
perturbative and non-perturbative domains in the QCD sum rule 
analysis. 

To the same order in the condensates with lowest dimension, 
the resulting time component of the isoscalar vector self-energy is 
\begin{equation} 
\Sigma_V^{(0)} = \frac{64 \pi^2}{3{\cal M}^2} \langle q^\dagger q \rangle_\rho 
= \frac{32 \pi^2}{{\cal M}^2} \rho \; .
\end{equation}  
It reflects the repulsive density-density correlations associated with the time component
of the quark baryon current, $\bar{q}\gamma^{\mu}q$. 
Note that around nuclear matter saturation density where $\rho_s \simeq \rho$, the ratio 
\begin{equation} 
\label{ratio} 
\frac{\Sigma_S^{(0)}}{\Sigma_V^{(0)}} = - \frac{\sigma_N}{4(m_u +m_d)} 
\frac{\rho_s}{\rho} 
\end{equation} 
is approximately equal to  $-1$ for typical values of 
the nucleon sigma term $\sigma_N$ and the current quark masses 
$m_{u,d}$  
(as an example, take $\sigma_N \simeq  50$ MeV and $m_u +m_d \simeq  
12$ MeV at a scale of 1 GeV). 

Identifying the free nucleon mass at $\rho=0$ with the Ioffe's formula 
Eq.~(\ref{Io}), 
$M = - \frac{8\pi^2}{{\cal M}^2} \langle \bar{q} q \rangle_0$, one finds 
\begin{equation} 
\label{back1} 
\Sigma_S^{(0)} (\rho) = M^*(\rho) -M = 
- \frac{\sigma_N M}{m_\pi^2 f_\pi^2} \rho_s 
\end{equation} 
and 
\begin{equation} 
\label{back2} 
\Sigma_V^{(0)} (\rho) = \frac{4 (m_u + m_d)M}{m_\pi^2f_\pi^2} \rho \; , 
\end{equation} 
with the quark masses $m_{u,d}$ taken at a renormalization 
scale $\mu \simeq {\cal M} \simeq 4\pi f_\pi \simeq 1$ GeV. 
Eq.~(\ref{back1}) implies (identifying $M$ with the free nucleon mass): 
\begin{equation} 
\Sigma_S^{(0)} \simeq -350~{\rm MeV}~\frac{\sigma_N} 
{50~{\rm MeV}} \,{\rho_s\over \rho_0} \; .
\end{equation}
Evidently, the leading-order in-medium change of the chiral condensate
is a source of a strong, attractive scalar field which acts
on the nucleon in such a way as to reduce its mass in nuclear matter
by more than $1/3$ of its vacuum value.

We note that the QCD sum rule estimates 
implied by Eqs.~(\ref{back1}), (\ref{back2}) and by the ratio Eq.~(\ref{ratio}) 
are not very accurate at a quantitative level. 
The leading-order Ioffe formula on which Eq.~(\ref{back1}) relies, should be 
corrected by contributions 
from condensates of higher dimension which are not well under control. 
The estimated error in the ratio $\Sigma_S^{(0)}/\Sigma_V^{(0)}\simeq -1$ 
is about 20\%, given the uncertainties in the values of $\sigma_N$ and $m_u+m_d$. 
Nevertheless, the constraint implied by Eq.~(\ref{ratio}) will be very useful as a 
starting point in constructing the relativistic energy density functional for finite nuclei.
%
\section{Relativistic Energy Density Functional Based on Chiral EFT}
\label{sec:FKVW}
%
Starting from the successful in-medium ChPT calculation of homogeneous 
nuclear matter explored in the previous section, we will develop a relativistic 
energy density functional (EDF) and 
apply it to studies of the structure of finite nuclei. The construction of this 
EDF is based on  the following conjectures  \cite{Fi.02,Fi.03,Fi.06}:
\begin{enumerate} 
\item 
The nuclear ground state is characterized by strong scalar and vector 
mean fields which have their origin in the in-medium changes of the 
scalar quark condensate and of the quark density. 
\item 
Nuclear binding and saturation arise primarily from chiral (pionic) 
fluctuations in combination with 
Pauli blocking effects and three-nucleon (3N) interactions, 
superimposed on the condensate background fields. 
\end{enumerate}
\subsection{The Nuclear Energy Density Functional}
\label{sec:NEDF}
From Sec. \ref{sec:DFT} where the basics of density functional theory (DFT) were 
introduced, we recall that the energy functional is commonly decomposed into three separate terms:
\begin{equation}
\label{KS}
F[\rho] = T_{kin}[\rho] + E_{H}[\rho] + E_{xc}[\rho] \; ,
\end{equation}
where $T_{kin}$ is the kinetic energy of 
the non-interacting N-particle system, $E_{H}$ is a Hartree
energy, and $E_{xc}$  denotes the exchange-correlation 
energy which, by definition, contains everything else. 
In the formulation of the relativistic EDF we assume that
the large scalar and vector mean fields
that have their origin in the in-medium changes of the 
chiral condensate and of the quark density, 
determine the Hartree energy functional $E_{H}[\rho]$, 
whereas the chiral (pionic) fluctuations including one- and
two-pion exchange with single and double 
virtual $\Delta$(1232)-isobar excitations plus Pauli blocking 
effects, are incorporated in the 
exchange-correlation energy functional $E_{xc}[\rho]$.

The density distribution and the energy of the nuclear ground state 
are obtained from self-consistent solutions of the relativistic 
generalizations of linear single-nucleon Kohn-Sham equations. 
In order to derive those equations it is useful to construct a
model with point-coupling NN interactions and density-dependent
vertices, designed in such way to reproduce the detailed
density dependence of the nucleon self-energies resulting from $E_H
[\rho] + E_{xc}[\rho]$. For a two component system of protons
and neutrons we start from a relativistic Lagrangian which includes 
isoscalar-scalar (S), isoscalar-vector (V),
isovector-scalar (TS) and isovector-vector (TV)
effective four-fermion interaction vertices with density-dependent 
coupling strengths. The minimal Lagrangian density reads:
\begin{equation}
\mathcal{L} = \mathcal{L}_{\rm free} + \mathcal{L}_{\rm int}^{(1)}
  + \mathcal{L}_{\rm int}^{(2)} + \mathcal{L}_{\rm coul} \; ,
\label{Lag}
\end{equation}
with the four terms defined as follows:
\begin{eqnarray}
\label{Lag2}
\mathcal{L}_{\rm free} & = &\bar{\psi}
   (i\gamma_{\mu}\partial^{\mu} -M_N)\psi \; ,\\
\label{Lag3}
\mathcal{L}_{\rm int}^{(1)} & = &
   - \frac{1}{2}~G_{S}(\hat{\rho}) (\bar{\psi}\psi)(\bar{\psi}\psi)
   -\frac{1}{2}~G_{V}(\hat{\rho})(\bar{\psi}\gamma_{\mu}\psi)
   (\bar{\psi}\gamma^{\mu}\psi) \nonumber\\
   & ~ & - \frac{1}{2}~G_{TS}(\hat{\rho})
   (\bar{\psi}\vect{\tau}\psi) \cdot (\bar{\psi} \vect{\tau} \psi)
   - \frac{1}{2}~G_{TV}(\hat{\rho})(\bar{\psi}\vect{\tau}
   \gamma_{\mu}\psi)\cdot (\bar{\psi}\vect{\tau} \gamma^{\mu}\psi) \; ,\\
\label{Lag4}
\mathcal{L}_{\rm int}^{(2)} & = & -\frac{1}{2}~D_{S} \partial_{\nu}(
   \bar{\psi}\psi) \partial^{\nu}(\bar{\psi}\psi)
\;, \\
\label{Lag5}
\mathcal{L}_{\rm em} & = &
   eA^{\mu}\bar{\psi}\frac{1+\tau_3}{2}\gamma_{\mu}\psi
   -\frac{1}{4} F_{\mu\nu}F^{\mu\nu} \; ,
\end{eqnarray}
where $\psi$ is the Dirac field of the nucleon with its two isospin 
components (p and n).
Vectors in isospin space are denoted by arrows.
In addition to the free nucleon Lagrangian $\mathcal{L}_{\rm free}$
and the interaction terms contained in
$\mathcal{L}_{\rm int}^{(1)}$, when applied to finite nuclei the model
must include the coupling $\mathcal{L}_{\rm em}$
of the protons to the electromagnetic field $A^\mu$ with $F_{\mu\nu} =
\partial_\mu A_\nu - \partial_\nu A_\mu$,
and a derivative (surface) term $\mathcal{L}_{\rm int}^{(2)}$. 
One could, of course, construct additional 
derivative terms in $\mathcal{L}_{\rm int}^{(2)}$,
further generalized to include density-dependent strength parameters. 
However, there appears to be no need in practical
applications to go beyond the simplest ansatz (\ref{Lag4})
with a constant $D_S$. Note that we do not introduce explicit spin-orbit terms. They emerge naturally from the Lorentz scalar and vector self-energies generated by the relativistic density functional. 

The variational principle $\delta\mathcal{L}/\delta \bar{\psi} =0$ applied to
the Lagrangian Eq.~(\ref{Lag}) leads to the self-consistent single-nucleon Dirac 
equations, the relativistic analogue of the (non-relativistic) 
Kohn-Sham equations. 
The nuclear dynamics produced by chiral (pionic) 
fluctuations in the medium is now encoded in the density dependence of the 
interaction vertices. In the framework of
relativistic density functional theory \cite{DG.90,Sp.92,Sc.95},
the density-dependent couplings are functions of the 4-current $j^\mu$:
\begin{equation}
j^\mu = \bar{\psi} \gamma^\mu \psi = \hat{\rho} u^{\mu} \; ,
\end{equation}
where $u^{\mu}$ is the 4-velocity defined as 
$(1-{\bm v}^2)^{-1/2}(1,{\bm v})$. We work in the rest-frame of the 
nuclear system with $\bm v=0$. 

The couplings $G_i(\hat{\rho})$ ($i=S,V,TS,TV$) are decomposed as follows: 
\begin{eqnarray}
\label{coupl}
G_i(\hat{\rho}) &=& G_i^{(0)} + G_i^{(\pi)} (\hat{\rho}) 
\quad ({\rm for}\,\, i=S,V)\nonumber \\
{\rm and} \quad G_i(\hat{\rho}) &=& G_i^{(\pi)} (\hat{\rho}) 
\quad \quad \quad\quad({\rm for}\,\,i=TS,TV)\; .
\end{eqnarray}
The density-independent parts $G_i^{(0)}$ arise from strong isoscalar 
scalar and vector background fields, whereas the density-dependent 
parts $G_i^{(\pi)} (\hat{\rho})$ are generated by one- and two-pion
exchange dynamics. It is assumed that only pionic processes
contribute to the isovector channels.

The relativistic density functional describing the 
ground-state energy of the system
can be re-written as a sum of four distinct terms:
\begin{equation}
\label{dft}
E_0[\hat{\rho}] = 
E_{\rm free} [\hat{\rho}] +
E_{\rm H} [\hat{\rho}] +
E_{\rm coul} [\hat{\rho}] +
E_{\rm \pi} [\hat{\rho}] \; ,
\end{equation}
with
\begin{eqnarray}
\label{e1}
E_{\rm free} [\hat{\rho}] & = & 
\int d^3r~\leftg \bar{\psi} [-i{\bm \gamma} \cdot {\bm \nabla} + M_N ] \psi 
\rightg  \; ,\\
\label{e2}
E_{\rm H} [\hat{\rho}] & = & \frac{1}{2}
\int d^3r~\{ \leftg G_S^{(0)}(\bar{\psi}\psi)^2 \rightg + \leftg
G_V^{(0)} (\bar{\psi} \gamma_\mu \psi)^2 \rightg \} \; , \\ 
\label{e3}
E_{\rm \pi} [\hat{\rho}] & = & \frac{1}{2}
\int d^3r~ \left\{ \leftg G_S^{(\pi)}(\hat{\rho})(\bar{\psi}\psi)^2 \rightg +
\leftg G_V^{(\pi)}(\hat{\rho}) (\bar{\psi} \gamma_\mu \psi)^2 \rightg  \right.
\nonumber \\ 
\label{e4}
& ~ & + \leftg G_{TS}^{(\pi)}(\hat{\rho}) 
(\bar{\psi}\vect{\tau} \psi)^2 \rightg 
+ \leftg G_{TV}^{(\pi)}(\hat{\rho})(\bar{\psi} \gamma_\mu \vect{\tau}
\psi)^2 \rightg  \nonumber \\
\label{e5}
& ~ & - \left. \leftg D_S^{(\pi)} 
[{\bm \nabla}(\bar{\psi}\psi)]^2 \rightg 
\right\} 
\; ,\\
\label{e6}
E_{coul} [\hat{\rho}] & = & \frac{1}{2}
\int d^3r~\leftg A^\mu e\bar{\psi} \frac{1+\tau_3}{2}\gamma_\mu \psi
\rightg \; ~~,
\end{eqnarray}
where $\rightg$ denotes the nuclear ground state.
Here $E_{\rm free}$ is the energy of the free (relativistic)
nucleons including their rest mass. $E_{\rm H}$ is a Hartree-type
contribution representing strong scalar and vector 
background condensate fields, and
$E_{\rm \pi}$ is the part of the energy generated by chiral $\pi N
\Delta$-dynamics, including a derivative (surface) term.
Minimization of the ground-state energy, represented in terms of a set of
auxiliary Dirac spinors ${\psi_k}$, leads to the relativistic analogue of the
Kohn-Sham equations. These single-nucleon Dirac equations are solved
self-consistently in the ``no-sea''
approximation which omits the explicit contribution 
of negative-energy solutions of the relativistic 
equations to the densities and currents. This means that vacuum polarization 
effects are not taken into account explicitly, but rather included in the adjustable 
parameters of the theory \cite{QHD1,QHD2}.

The expressions for the isoscalar and isovector four-currents and 
scalar densities read: 
\begin{eqnarray}
j_\mu & =
 \sum_{k=1}^N \bar{\psi}_k \gamma_\mu \psi_k \; , \quad\quad\quad 
\vect{j}_\mu & =  
 \sum_{k=1}^N \bar{\psi}_k \gamma_\mu \vect{\tau} \psi_k \; ,\\
\rho_S & = 
 \sum_{k=1}^N \bar{\psi}_k \psi_k \; , \quad\quad\quad
\vect{\rho}_S & = 
 \sum_{k=1}^N \bar{\psi}_k \vect{\tau} \psi_k \; ,
\end{eqnarray}
where $\psi_k$ are Dirac spinors and
the sum runs over occupied positive-energy single-nucleon states.
If we only consider systems with time-reversal
symmetry in the ground-state, i.e. even-even nuclei, 
the space components of all 
currents vanish (${\bf j}=0$) and, because of
charge conservation, only the third component of
isospin ($\tau_3 = -1$ for neutrons and $ \tau_3 = +1 $ for protons)
contributes. The relevant combinations of densities are:
\begin{eqnarray}
   \rho&=  \leftg \bar{\psi}\gamma^{0} \psi \rightg
    = \rho^p + \rho^n \; , \quad\quad\quad
   \rho_{3} &=  \leftg \bar{\psi} \tau_3 \gamma^{0} 
   \psi \rightg
   = \rho^p - \rho^n  \; ,\\
   \rho_S &=  \leftg \bar{\psi} \psi \rightg 
    = \rho_S^p + \rho_S^n \; , \quad\quad\quad 
   \rho_{S3} &=  \leftg \bar{\psi} \tau_3 \psi \rightg 
    = \rho_S^p - \rho_S^n \; .
\end{eqnarray}
Minimization with respect to $\bar{\psi}_k$ gives the single-nucleon Dirac equation:
\begin{equation}
\label{mean-field}
[-i{\bm \gamma}\cdot {\bm \nabla} + M_N
+ \gamma_0 \Sigma_V + \gamma_0 \tau_3 \Sigma_{TV}
+\gamma_0 \Sigma_R + \Sigma_S + \tau_3 \Sigma_{TS}
] \psi_k = \epsilon_k \psi_k \; ,
\end{equation}
with the self-energies:
\begin{eqnarray}
\label{self1}
\Sigma_V & = & [G_V^{(0)} + G_V^{(\pi)}(\rho)] \rho 
+ eA^0\frac{1+\tau_3}{2}\; ,\\
\label{self2}
\Sigma_{TV} & = & G_{TV}^{(\pi)}(\rho)\, \rho_{3} \; ,\\
\label{self3}
\Sigma_{S}   & = & [G_S^{(0)} + G_S^{(\pi)}(\rho)] \rho_S + D_S^{(\pi)}
{\bm \nabla}^2 \rho_S \; ,\\
\label{self4}
\Sigma_{TS} & = & G_{TS}^{(\pi)}(\rho)\, \rho_{S3} \; ,\\
\label{self5}
\Sigma_R & = & \frac{1}{2} \left\{ \frac{\partial G_V^{(\pi)}(\rho)
}{\partial \rho}
\rho^2 +   \frac{\partial G_S^{(\pi)}(\rho)
}{\partial \rho} \rho_S^2\, + \right. \nonumber \\
 & ~ & \left. \hspace{3cm} \frac{\partial G_{TV}^{(\pi)}(\rho)
}{\partial \rho}
\rho_{3}^2 + \frac{\partial G_{TS}^{(\pi)}(\rho)
}{\partial \rho}
\rho_{S3}^2 \right\} \; ,
\end{eqnarray}
where $A^0 ({\bf r})$ in Eq.~(\ref{self1}) is the Coulomb potential.
In addition to the usual contributions from the time components of the 
vector self-energies and the scalar potentials, we must also 
include the ``rearrangement'' 
term, $\Sigma_R$, arising from the variation of the vertex 
functionals with respect to the nucleon fields
in the density operator $\hat{\rho}$. 
For a Lagrangian with density-dependent couplings, the inclusion
of the rearrangement self-energies
is essential in order to guarantee energy-momentum conservation and 
thermodynamical consistency ${\rho^2 \frac{\partial}{\partial \rho}
\left( \frac{\varepsilon}{\rho} \right) = \frac{1}{3}
\sum_{i=1}^3 T^{ii} }$ 
(i.e. for the pressure equation derived from 
the thermodynamic definition and from the 
energy-momentum tensor). Using the single-nucleon Dirac equation and
performing an integration by parts, the ground-state energy of a 
nucleus with A nucleons reads:
\begin{eqnarray}
  E_0 = 
   \sum\limits_{k=1}^{A} \epsilon_k & ~ & 
   - \frac{1}{2} \int d^3r \left\{
   ~[G_S^{(0)} + G_S^{(\pi)}(\rho)] \,\rho_S^2 + 
    G_{TS}^{(\pi)}(\rho) \,\rho_{S3}^2 +
   [G_V^{(0)} + G_V^{(\pi)}(\rho)]\, \rho^2 +  \right. \nonumber\\
     & ~ & G_{TV}^{(\pi)}(\rho)\, \rho_{3}^2 + 
    \frac{\partial G_S^{(\pi)}(\rho)}{\partial \rho}\,\rho_S^2\, \rho +
   \frac{\partial G_{TS}^{(\pi)}(\rho)}{\partial \rho} \,\rho^2_{S3}\, \rho +
   \nonumber\\ 
 & ~ &
   \frac{\partial G_V^{(\pi)}(\rho)}{\partial \rho}\, \rho^3 +
   \frac{\partial G_{TV}^{(\pi)}(\rho)}{\partial \rho} \,\rho^2_{3}\, 
   \rho +
 \left. e\,\rho_{ch}\, A^0 + D_S^{(\pi)} \rho_S \nabla^2 \rho_S 
   \right\} \; ,  
\end{eqnarray}
where $\epsilon_k$ denotes the single-nucleon Kohn-Sham energies.
%
\subsection{Low-energy QCD Constraints}
\label{sec:Link_QCD}
%
In Sec.~\ref{sec:NM-QCDSR} we have shown how the in-medium QCD sum rules 
relate the leading changes of the scalar quark 
condensate and of the quark density at finite baryon density, 
with the scalar and vector self-energies
of a nucleon in the nuclear medium. Comparing 
Eqs. (\ref{self1}) and (\ref{self3}) for the isoscalar vector and 
scalar potentials of the single-nucleon Dirac equations, with the Eqs.
(\ref{back1}) and (\ref{back2}) for the condensate background self-energies,
respectively, the following estimates are obtained for the 
couplings of the nucleon to the background fields (the Hartree terms in the 
energy functional):
\begin{equation}
G_S^{(0)} = - \frac{\sigma_N M_N}{m_\pi^2 f_\pi^2} 
 \simeq  - 11~{\rm fm}^{2}~\left[ \frac{\sigma_N}{50~{\rm MeV}}\right] \; ,
\end{equation}
and
\begin{equation}
G_V^{(0)} = \frac{4(m_u + m_d)M_N}{m_\pi^2 f_\pi^2} 
 \simeq  11~{\rm fm}^{2}~\left[ \frac{4(m_u + m_d)}{50~{\rm MeV}}\right]\; .
\end{equation}
\bigskip

The many-body effects represented by the exchange-correlation 
density functional are approximated by chiral 
$\pi N \Delta$-dynamics, including Pauli blocking effects. 
In the simplest DFT approach, the exchange-correlation 
energy for a finite system is determined 
in the local density approximation (LDA)
from the exchange-correlation functional of the corresponding 
infinite homogeneous system,
replacing the constant density $\rho$ by the
local density $\rho ({\bf r})$ of the actual inhomogeneous system.
In our case the exchange-correlation terms of the nuclear density
functional are determined within LDA by equating the 
corresponding self-energies in the single-nucleon Dirac equation 
(\ref{mean-field}), with those arising from the in-medium 
chiral perturbation theory calculation of 
$\pi N \Delta$-dynamics in homogeneous isospin symmetric and
asymmetric nuclear matter (cf. Sec.~\ref{sec:NM}):
\begin{eqnarray}
\label{sum}
U(p = k_f,\rho) & = & \Sigma_S^{\rm ChPT} (k_f,\rho) + 
\Sigma_V^{\rm ChPT} (k_f,\rho) \nonumber \\
- U_I(p = k_f,\rho) \delta & = & \Sigma_{TS}^{\rm ChPT} (k_f,\rho) +
\Sigma_{TV}^{\rm ChPT} (k_f,\rho)\; ,
\end{eqnarray}
where $U(p,k_f)$ and $U_I(p,k_f)$ are the isoscalar and isovector 
momentum and density-dependent single-particle potentials, respectively, 
and  $\delta = ({\rho^n - \rho^p})/({\rho^n + \rho^p})$.
In order to determine the density-dependent couplings of the 
exchange-correlation pieces generated by the point-coupling model, 
a polynomial fit up to order $k_f^6$ is performed for the 
ChPT self-energies, and they are re-expressed in terms of the baryon
density $\rho =  \rho^p + \rho^n$ 
and the isovector density $\rho_3 = \rho^p - \rho^n$:
\begin{eqnarray}
  \Sigma_S^{\rm ChPT}(k_f,\rho) & = & (c_{s1} + c_{s2}\rho^{\frac{1}{3}}
  + c_{s3}\rho^{\frac{2}{3}}+ c_{s4} \rho) \,\rho
  \; ,\label{prho1} \\
  \Sigma_V^{\rm ChPT}(k_f,\rho) & = & (c_{v1} + c_{v2}\rho^{\frac{1}{3}}
  + c_{v3}\rho^{\frac{2}{3}}+ c_{v4} \rho)\, \rho
  \; ,\label{prho2}\\
  \Sigma_{TS}^{\rm ChPT}(k_f,\rho) &=& (c_{ts1} + c_{ts2}\rho^{\frac{1}{3}}
  + c_{ts3}\rho^{\frac{2}{3}}+ c_{ts4} \rho )\, \rho_{3} 
  \; ,\label{prho3}\\
  \Sigma_{TV}^{\rm ChPT}(k_f,\rho) &=& (c_{tv1} + c_{tv2}\rho^{\frac{1}{3}}
  + c_{tv3}\rho^{\frac{2}{3}}+ c_{tv4} \rho)\, \rho_{3}
  \label{prho4}\; ,
\end{eqnarray}
Next these self-energies are identified with the corresponding 
contributions to the 
potentials Eqs. (\ref{self1})-(\ref{self5}) in the point-coupling single-nucleon 
Dirac equation. 
The resulting expressions for the density-dependent couplings
of the pionic fluctuation terms read (
small differences between $\rho$ and $\rho_S$ at nuclear matter densities
are neglected here):
\begin{eqnarray}
\label{GS}
  G_S^{(\pi)} (\rho)& = & c_{s1} + c_{s2} \rho^{\frac{1}{3}}
  + c_{s3} \rho^{\frac{2}{3}} + c_{s4} \rho \; ,\\
\label{GV}
  G_V^{(\pi)} (\rho)& = & {\bf \bar{c}_{v1}} + 
{\bf \bar{c}_{v2}} \rho^{\frac{1}{3}}
  + {\bf \bar{c}_{v3}} \rho^{\frac{2}{3}} + {\bf \bar{c}_{v4}} \rho \; ,\\
\label{GTS}
  G_{TS}^{(\pi)} (\rho)& = & c_{ts1} + c_{ts2} \rho^{\frac{1}{3}}
  + c_{ts3} \rho^{\frac{2}{3}} + c_{ts4} \rho \; ,\\
\label{GTV}
  G_{TV}^{(\pi)} (\rho)& = & c_{tv1} + c_{tv2} \rho^{\frac{1}{3}}
  + c_{tv3} \rho^{\frac{2}{3}} + c_{tV4} \rho \; ,
\end{eqnarray}
where the inclusion of the rearrangement term $\Sigma_R$ redefines
the isoscalar-vector coefficients: ${\bf \bar{c}_{v1}}  =  c_{v1}$, 
${\bf \bar{c}_{v2}}  =  {1}/{7}(6c_{v2}-c_{s2})$, ${\bf \bar{c}_{v3}}  =  
{1}/{4}(3c_{v3}-c_{s3})$, and ${\bf \bar{c}_{v4}}  =  
{1}/{3}(2c_{v4}-c_{s4})$ \cite{Fi.03}.

The coefficient $D_{S}^{(\pi)}$ of the derivative term in the 
equivalent point-coupling model (Eq.~(\ref{self3}))
can be determined from ChPT calculations for inhomogeneous nuclear matter.
The isoscalar nuclear energy density emerging 
from chiral pion-nucleon dynamics has the form \cite{Fri.04}:
\begin{eqnarray} 
\label{Skyrme}
{\mathcal E}(\rho, {\bm \nabla }\rho)& = & \rho\,\bar E(k_f) 
+ ({\bm \nabla }\rho)^2 \, F_\nabla(k_f) + \ldots \; . 
\end{eqnarray} 
$F_\nabla$ can be approximated by a constant in the relevant region of 
nuclear densities. The following relation between $F_\nabla$ and the 
derivative term of the point-coupling model (see Eqs.~(\ref{Lag4},\ref{e5})) is then 
valid:
\begin{equation}
- 2 {F_\nabla} = D_S^{(\pi)} \; .
\label{ds} 
\end{equation} 
The inclusion of derivative terms in the model Lagrangian and the 
determination of its strength parameters from ChPT 
calculations in inhomogeneous matter actually goes beyond the 
local density approximation. The term Eq.~(\ref{Lag4}), 
with the strength parameter given by Eq.~(\ref{ds}), represents a
second-order gradient correction to the LDA, i.e. the next-to-leading term 
in the gradient expansion of the exchange-correlation energy calculated 
by in-medium chiral perturbation theory. 

While bulk properties of infinite nuclear matter are useful for 
orientation, the large amount of nuclear observables  
provides a far more accurate data base that 
permits a fine tuning of the global parameters.
The total number of adjustable parameters of 
the point-coupling model is seven, four of which are related to 
contact (counter) terms that appear in the 
ChPT treatment of nuclear matter. One parameter 
fixes a surface (derivative) term, and two more represent the strengths of 
scalar and vector Hartree fields.  The values of the parameters
are adjusted simultaneously to properties of nuclear matter and 
to binding energies, charge radii and
differences between neutron and proton radii of spherical nuclei, 
starting from the estimates  for the couplings 
of the condensate background fields (Hartree term), and 
the constants in the expressions  
for the self-energies arising from chiral
$\pi N \Delta$-dynamics (exchange-correlation term). 
The resulting optimal parameter set (FKVW) \cite{Fi.06} is 
remarkably close to the anticipated QCD sum rule and ChPT values,
with the exception of the two
constants associated with three-body correlations, for which 
the fit to nuclear data systematically requires an attractive shift 
as compared to the ChPT calculation \cite{Fri.04}.
%
\subsection{Nuclear Ground-State Properties}
\label{sec:Finite_Nuclei}
%
The effective global FKVW interaction has been tested in 
self-consistent calculations of ground-state observables for spherical 
and deformed medium-heavy and heavy nuclei. The calculations, including
open-shell nuclei, are performed in the framework of the relativistic 
Hartree-Bogoliubov (RHB) model, a relativistic
extension of the conventional Hartree-Fock-Bogoliubov method,
that provides a basis for a consistent microscopic 
description of ground-state properties, low-energy excited states, 
small-amplitude vibrations, and 
reliable extrapolations toward the drip lines \cite{VALR.05}. 
The new microscopic FKVW interaction \cite{Fi.06} has been employed
in the particle-hole channel, in comparison with one of the 
most successful phenomenological effective meson-exchange 
relativistic mean-field interactions: DD-ME1  \cite{Nik.02}.
Pairing effects in nuclei are restricted to a narrow window of a few MeV 
around the Fermi level. Their scale
is well separated from the scale of binding energies which
are in the range of several hundred to thousand MeV, and 
thus pairing can be treated as a non-relativistic phenomenon.
In most applications of the RHB model the pairing part of the well 
known and successful Gogny force~\cite{BGG.84} has been used
in the particle-particle channel, and the same interaction is used in the 
illustrative examples included in this section. 
\begin{figure}[h]
\begin{center}
\vspace{0.5 cm}
\includegraphics[scale=0.45,angle=0]{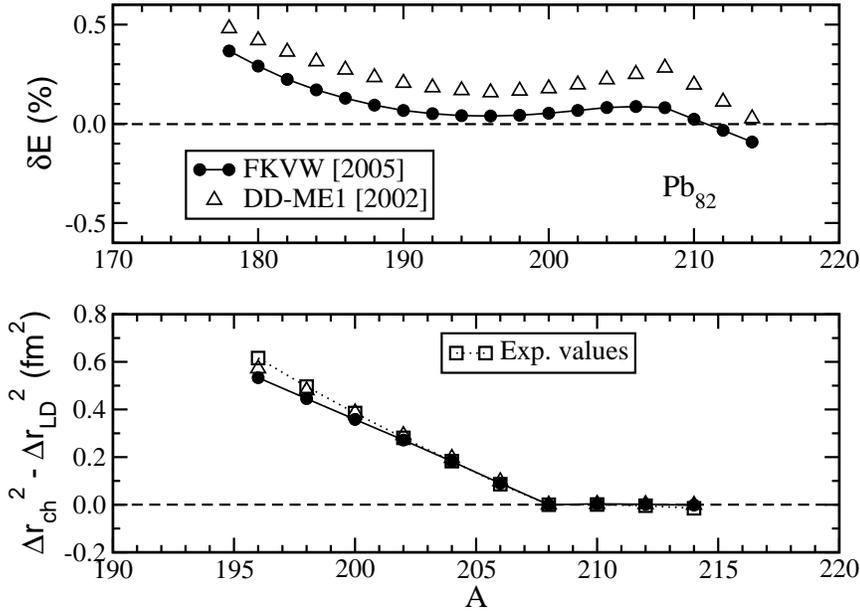}
\end{center}
\caption{\label{lead}
The deviations (in percent) of the calculated binding 
energies from the experimental values (upper panel)~\cite{Audi2003}, 
and the calculated charge
isotope shifts in comparison with data~\cite{Rad}, 
for the chain of even-A Pb isotopes. 
The charge isotope shifts are defined: 
$\Delta r_{ch}^2 = r_{ch}^2(A) - r_{ch}^2(^{208}Pb)$ and 
$ \Delta r_{LD}^2 = r_{LD}^2(A) - r_{LD}^2(^{208}Pb)$, where
the liquid-drop estimate is $r^2_{LD}(A) = {3\over 5}r_0^2A^{2/3}$.
}
\end{figure}

The isotopic dependence of the 
deviations (in percent) between the calculated binding 
energies and the experimental values for even-A Pb nuclei, 
is plotted in the upper panel of Fig.~\ref{lead}. It is 
interesting to note that, although DD-ME1 and FKVW represent 
different physical models, they display a 
similar mass dependence of the calculated binding energies
for the Pb isotopic chain. On a quantitative level the FKVW 
interaction produces better results, with the absolute deviations
of the calculated masses below 0.1 \% for $A \geq 190$. In lighter
Pb isotopes one expects that the observed shape coexistence phenomena 
will have a pronounced effect on the measured masses. 
Because of the intrinsic isospin dependence of the effective
single-nucleon spin-orbit potential, relativistic mean-field
models naturally reproduce the anomalous charge isotope
shifts. The well known example
of the anomalous kink in the charge isotope shifts of
Pb isotopes is illustrated in the lower panel of Fig.~\ref{lead}. 
The results of RHB calculations with the DD-ME1 and FKVW effective 
interactions are shown in comparison with experimental values.
Both interactions reproduce in detail the A-dependence
of the isotope shifts and the kink at $^{208}$Pb.

\begin{figure}[h]
\begin{center}
\vspace{1 cm}
\includegraphics[scale=0.45,angle=0]{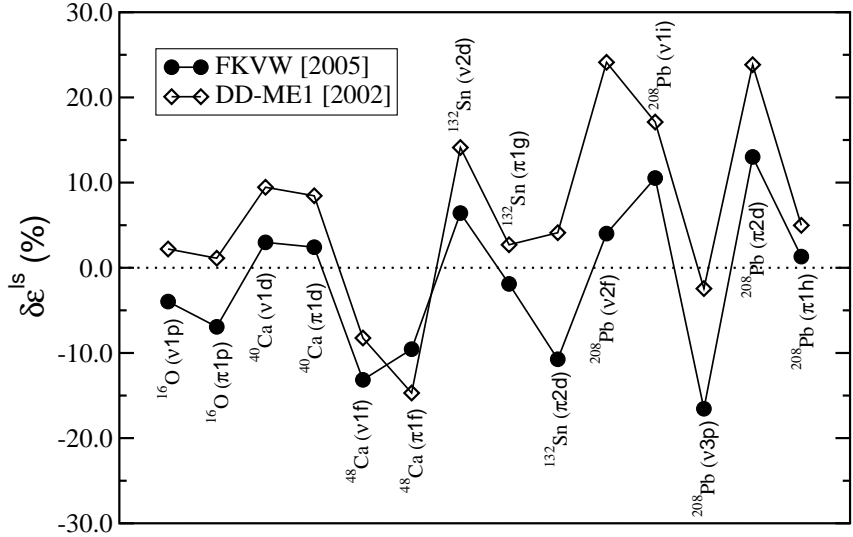}
\end{center}
\caption{\label{spin_orbit}
The deviations (in percent) between the theoretical and experimental 
values \cite{NUDAT}, of the energy spacings between spin-orbit partner-states in 
doubly closed-shell nuclei. }
\end{figure}

One of the principal advantages of using the 
relativistic framework lies in the fact
that the effective single-nucleon spin-orbit potential 
arises naturally from the Dirac equation. 
The single-nucleon potential does not include any adjustable parameter for 
the spin-orbit interaction. In the FKVW model, in particular, 
the large effective spin-orbit potential in finite nuclei is generated
by the strong scalar and vector condensate background 
fields of about equal magnitude and
opposite sign, induced by changes of the QCD vacuum in the presence
of baryonic matter.  Fig. \ref{spin_orbit} displays
the deviations (in percent) between the calculated and experimental 
values of the energy spacings between spin-orbit partner-states in
a series of doubly closed-shell nuclei. 
The theoretical spin-orbit splittings 
have been calculated with the FKVW and DD-ME1 interactions. 
For the phenomenological DD-ME1 interaction 
the large scalar and vector nucleon self-energies which generate 
the spin-orbit potential, arise from the exchange of ``sigma'' and
``omega'' mesons with adjustable strength parameters.
One notices that, even though the values calculated with DD-ME1 are already
in very good agreement with experimental data, a further improvement
is obtained with the FKVW interaction. This remarkable agreement  
indicates that the initial estimates  for the condensate background 
couplings have been more realistic than anticipated,  
considering the uncertainties of lowest-order in-medium QCD sum rules. 
\vspace{0.1 cm}
\begin{figure}[h]
\begin{center}
\vspace{0.5 cm}
\includegraphics[scale=0.58,angle=0]{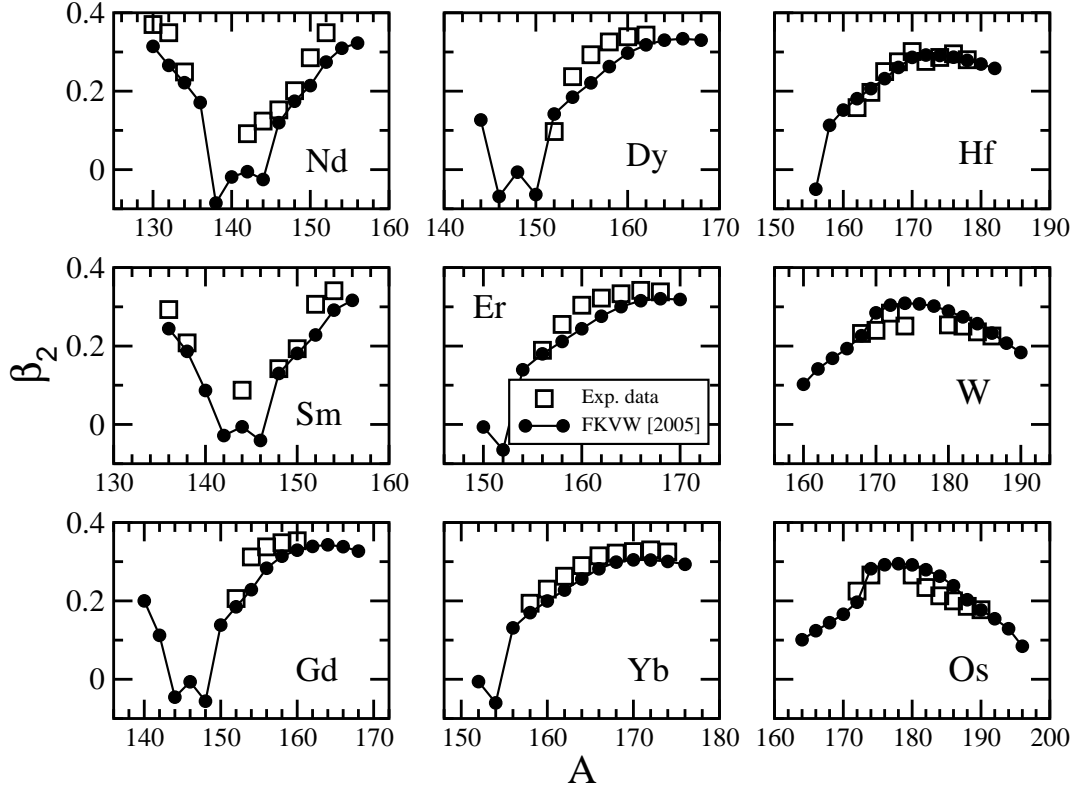}
\end{center}
\caption{\label{Beta_def}
Comparison between the RHB model (FKVW interaction plus Gogny pairing) 
predictions for the ground-state
quadrupole deformation parameters of the Nd, Sm, Gd, 
Dy, Er, Yb, Hf, W, and Os
isotopes, and experimental values \cite{RNT.01}.
}
\end{figure}

Deformed nuclei with $N > Z$ present
further important tests for nuclear structure models. 
Ground-state properties, in particular, are sensitive 
to the isovector channel of the effective interaction, 
to the spin-orbit term of the effective 
single-nucleon potentials and to the effective mass.
The nuclear density functional constrained by low-energy 
QCD has been tested in the region $60\le Z \le80$. 
Predictions of the RHB calculations for the total binding energies, 
charge radii and ground-states quadrupole deformations of 
even-Z isotopic chains have been compared with available data. 
With the FKVW effective interaction in the particle-hole
channel, and pairing correlations described by the 
finite range Gogny D1S interaction, very good agreement with 
experimental values has been found not only for the binding energies and 
charge radii over the entire region of deformed nuclei, but excellent results 
have also been obtained for the ground-state quadrupole deformations. 
The level of agreement with data is illustrated in Fig.~\ref{Beta_def}
where, for the chains of 
Nd, Sm, Gd, Dy, Er, Yb, Hf, W, and Os isotopes,
the calculated ground-state quadrupole deformation parameters $\beta_2$, 
proportional to the expectation value of the quadrupole 
operator $\langle \phi_0| 3 z^2 - r^2 |\phi_0 \rangle$,
are displayed in comparison with the empirical data extracted 
from $B(E2)$ transitions.
We notice that the RHB results reproduce not only the 
global trend of the data but also the saturation of quadrupole deformations
for heavier isotopes.
%
\section{Concluding Remarks}
\label{sec:Conclusion}
%
We have reviewed a novel microscopic framework of nuclear 
energy density functionals, which synthesizes effective field theory 
methods and principles of density functional theory, to establish a
direct link between low-energy QCD and nuclear structure.
At low-energies characteristic for nuclear binding, 
QCD is realized as a theory of pions coupled to nucleons. 
The basic concept of a low-energy EFT is the separation of scales:
the long-range physics is treated explicitly (e.g. pion exchange) 
and short-distance interactions, that cannot be resolved at low-energy,
are replaced by contact terms. The EFT building of a universal
microscopic energy density functional allows error estimates to be
made, and it also provides a power counting scheme which separates
long- and short-distance dynamics.

A relativistic nuclear energy density functional (EDF) has been introduced
and constrained by two closely related features of QCD in the 
low-energy limit: a) in-medium changes of vacuum condensates, and 
b) spontaneous chiral symmetry breaking. 
The changes of the chiral (quark) condensate and quark density 
in the presence of baryonic matter are sources of strong (attractive) 
scalar and (repulsive) vector fields experienced by nucleons in the 
nucleus. These fields produce Hartree mean-field nucleon potentials,
and are at the origin of the large energy spacings between 
spin-orbit partner states in nuclei. 

The spontaneously broken chiral symmetry in QCD introduces pions 
as Goldstone bosons, and determines the pion-nucleon couplings. 
Starting from in-medium chiral perturbation theory 
calculation of homogeneous nuclear matter, the
exchange-correlation part of the EDF is deduced 
from the long- and intermediate-range interactions generated by one- 
and two-pion exchange processes, with explicit inclusion
of $\Delta(1232)$ degrees of freedom. Regularization dependent 
contributions to the energy density, calculated 
at three-loop level, are absorbed in contact interactions
with constants representing unresolved short-distance dynamics. 

The EDF framework is realized as a relativistic point-coupling model with 
density-dependent interaction vertices. The solution of the system 
of self-consistent single-nucleon Dirac equations 
(relativistic analogues of  Kohn-Sham equations) determines the nucleon 
densities which enter the energy functional. 
The construction of the density functional involves an expansion of 
nucleon self-energies in powers of the Fermi momentum up to and including 
terms of order $k_f^6$, or equivalently, $O(\rho^2)$ in the proton and 
neutron densities. The exchange-correlation energy functional, 
determined by chiral  $\pi N\Delta$-dynamics in nuclear matter, 
is used in Kohn-Sham calculations of finite nuclei by employing 
a second-order gradient correction to the local density approximation.

Excellent results have been obtained in studies of ground-state properties 
(binding energies, charge form factors, radii of proton and neutron 
distributions, deformations, spin-orbit splittings) of spherical and 
deformed nuclei all over the chart of nuclides. The microscopic EDF 
reproduces data on the same level of comparison as the 
best fully phenomenological (non-relativistic and relativistic) 
self-consistent mean-field models. These results 
demonstrate that chiral EFT establishes a
consistent microscopic framework in which both the isoscalar and isovector 
channels of a universal nuclear energy density functional can be formulated. 
Guided by the principles of density functional theory and based on chiral 
EFT, the construction of the nuclear EDF provides a quantitative 
{\em ab initio} description of the complex nuclear many-body problem 
starting from the fundamental theory of strong interactions. 

\bigskip
\noindent
{\bf Acknowledgements}
\bigskip \\
\noindent
I would like to thank  my collaborators Paolo Finelli, Norbert 
Kaiser and Wolfram Weise, for their contribution to the subject 
reviewed in these lectures.
%
%

\end{document}